\pdfoutput=1

\documentclass[iop, apj]{emulateapj}

\usepackage{xspace}
\usepackage{amsmath}
\usepackage{framed} 
\usepackage{txfonts}
\usepackage{epstopdf} 
\usepackage{color}
\usepackage{rotating}
\usepackage{natbib}
\usepackage{ulem}
\usepackage{xspace}

\usepackage{hyperref}
\hypersetup{
  colorlinks   = true, 
  urlcolor     = black, 
  linkcolor    = blue, 
  citecolor   = blue 
}

\journalinfo{The Astrophysical Journal, {\rm 779:159 (11pp), 2013 December 20}}
\submitted{Received 2013 June 13; accepted 2013 October 21; published 2013 December 3}

\newcommand{\kpch}{\>{h^{-1}{\rm kpc}}}

\newcommand{\gpccubed}{\>({\rm {Gpc}/h)}^3}

\newcommand{\msunh}{\>h^{-1}\rm M_\odot}

\newcommand{\kmsmpc}{\>{\rm km}\,{\rm s}^{-1}\,{\rm Mpc}^{-1}}

\def\LCDM{$\Lambda$CDM\ }

\def\mvir{M_{\rm vir}}
\def\rvir{R_{\rm vir}}
\def\gcm3{\mathrm{g} / \mathrm{cm}^3}

\def\gtsima{$\; \buildrel > \over \sim \;$}
\def\ltsima{$\; \buildrel < \over \sim \;$}
\def\prosima{$\; \buildrel \propto \over \sim \;$}
\def\gsim{\lower.7ex\hbox{\gtsima}}
\def\lsim{\lower.7ex\hbox{\ltsima}}
\def\simgt{\lower.7ex\hbox{\gtsima}}
\def\simlt{\lower.7ex\hbox{\ltsima}}
\def\simpr{\lower.7ex\hbox{\prosima}}

\def\rmb{{\rm b}}

\def\rmm{{\rm m}}

\def\rms{{\rm s}}

\def\ms{$M-\sigma$\xspace}
\def\msr{$M-\sigma$ relation\xspace}

\def\sigr{\sigma(r)}
\def\sigsr{\sigma(<r)}
\def\sgmr{\sigsr / [GM(<r)/r]^{1/2}}
\def\sp{\sigma^\prime}

\def\rdelta{R_{\Delta}}
\def\mdelta{M_{\Delta}}
\def\rs{r_{\rm s}}
\def\fpe{f_{\rm PE}}

\def\rhot{\rho_{\Delta}}
\def\rhoc{\rho_{\rm c}}

\def\cvir{c_{\rm vir}}

\def\mtom{M_{\rm 200m}}
\def\rtom{R_{\rm 200m}}

\def\mtoc{M_{\rm 200c}}
\def\rtoc{R_{\rm 200c}}

\def\mfoc{M_{\rm 500c}}
\def\rfoc{R_{\rm 500c}}

\def\mtfc{M_{\rm 2500c}}
\def\rtfc{R_{\rm 2500c}}

\shorttitle{Evolution of scaling relations}
\shortauthors{Diemer, Kravtsov \& More}

\begin{document}


\def\figdir{.}
\def\figext{pdf}


\title{On the evolution of cluster scaling relations}

\author{Benedikt Diemer \altaffilmark{1,2}, Andrey V. Kravtsov\altaffilmark{1,2,3} and Surhud More \altaffilmark{4}}

\affil{
$^1$ Department of Astronomy \& Astrophysics, The University of Chicago, Chicago, IL 60637, USA; bdiemer@oddjob.uchicago.edu \\
$^2$ Kavli Institute for Cosmological Physics, The University of Chicago, Chicago, IL 60637, USA \\
$^3$ Enrico Fermi Institute, The University of Chicago, Chicago, IL 60637, USA \\
$^4$ Kavli Institute for the Physics and Mathematics of the Universe (WPI), The University of Tokyo, 5-1-5 Kashiwanoha, Kashiwa-shi, Chiba, 277-8583, Japan
}


\begin{abstract}

Understanding the evolution of scaling relations between the observable properties of clusters and their total mass is key to realizing their potential as cosmological probes. In this study, we investigate whether the evolution of cluster scaling relations is affected by the spurious evolution of mass caused by the evolving reference density with respect to which halo masses are defined (pseudo-evolution). We use the relation between mass, $M$, and velocity dispersion, $\sigma$, as a test case, and show that the deviation from the \msr of cluster-sized halos caused by pseudo-evolution is smaller than 10\% for a wide range of mass definitions. The reason for this small impact is a tight relation between the velocity dispersion and mass profiles, $\sigma(<r) = {\rm const} \times  \sqrt{GM(<r) / r}$, which holds across a wide range of radii. We show that such a relation is generically expected for a variety of density profiles, as long as halos are in approximate Jeans equilibrium. Thus, as the outer ``virial'' radius used to define the halo mass, $R$, increases due to pseudo-evolution, halos approximately preserve their \msr. This result highlights the fact that tight scaling relations are the result of tight equilibrium relations between radial profiles of physical quantities. We find exceptions at very small and very large radii, where the profiles deviate from the relations they exhibit at intermediate radii. We discuss the implications of these results for other cluster scaling relations, and argue that pseudo-evolution should have a small effect on most scaling relations, except for those that involve the stellar masses of galaxies. In particular, we show that the relation between stellar-mass fraction and total mass is affected by pseudo-evolution and is largely shaped by it for halo masses $\lesssim 10^{14}\ M_{\odot}$.
\end{abstract}



\section{Introduction}
\label{sec:intro}
 
Clusters of galaxies are  excellent laboratories for studying galaxy formation because their deep potential wells allow them to retain most of the associated baryons, the bulk of which are observable \citep[see, e.g.,][hereafter KB12, for a review]{kravtsov_12_cluster_review}. Clusters are also used as cosmological probes because their abundance constrains the exponential high-mass end of the halo mass function and thus important cosmological parameters \citep[e.g.,][]{eke_96_clustercosmo,  vikhlinin_09_xray, rozo_10_clustercosmo, allen_11_clusterreview}. The total halo masses of clusters are either observed directly through weak lensing \citep[e.g.,][]{mellier_99_lensingreview, smith_03_lensing, cypriano_04_lensing}, or reconstructed from proxy observables such as X-ray luminosity, X-ray temperature, or a combination thereof \citep[e.g.,][]{evrard_96_scaling,bryan_98_virial,kravtsov_06_mass_indicator,vikhlinin_06_clusters, bohringer_07_rexcess}, velocity dispersion \citep{yahil_77, becker_07} or the Sunyaev-Zel'dovich effect \citep[SZ,][]{sunyaev_72, benson_13_spt_xray}.

Observational and theoretical studies show that the scaling relations between such ``observables'' and total halo mass follow a power law form (e.g., KB12). The evolution of these scaling relations is often described within a framework of self-similar collapse \citep{kaiser_86_clusters, kaiser_91}, based on the spherical collapse of top hat fluctuations \citep{gunn_72_sphericalcollapse}. In this picture, an entire perturbation is expected to collapse at a well-defined epoch and have a well-defined radius that encloses all of the collapsed mass and corresponds to a virial overdensity predicted by the model. For realistic perturbation profiles, however, the collapse is extended in time, and the resulting density profile is extended in radius \citep{gott_75, gunn_77, fillmore_84, bertschinger_85, lithwick_11}. In particular, the overdensity profile at the center of the perturbation is shallow, leading to the inner shells collapsing at about the same time, and thus a fast growth in mass. In turn, the overdensity profile is steeper at larger radii, and the outer shells collapse slower, leading to a lower mass-growth rate and an outer profile with $\rho(r) \propto r^{-3}$ \citep{dalal_10, lithwick_11}. These fast and slow mass-growth regimes are generically observed in simulations \citep{wechsler_02_halo_assembly, zhao_03_mah, zhao_09_mah}.

In practice, i.e., for the commonly used mass definitions, halo mass can grow as a result of mergers, accretion, and changes of the halo boundary because of its definition. For example, in the spherical overdensity definition \citep[e.g.,][]{cole_96_halostructure}, the outer halo boundary is defined to enclose a certain overdensity,
\begin{equation}
\label{eq:sodef}
M_{\Delta} = \frac{4 \pi}{3} \rhot(z) R_{\Delta}^3 \,.
\end{equation}
The enclosed average density $\rhot$ is defined as a multiple of the critical or mean matter density of the universe at the redshift in question,
\begin{equation}
\label{eq:rhot}
\rhot(z) \equiv \Delta(z) \rho_{\rm ref}(z)
\end{equation}
where $\Delta$ may either be a constant or a function of redshift. The decrease of $\rhot$ with time leads to an increase of the halo radius and the corresponding enclosed mass. We call this increase ``pseudo-evolution'' because it depends on the mass definition and results in a mass change even when the density distribution around a halo does not change at all \citep[][hereafter D13]{diemer_13_pe}.

Although the pseudo-evolution of halo mass is non-zero at all stages of halo evolution as long as the reference density evolves, its contribution is a small fraction of the total mass growth for halos that accrete matter at a high rate (the fast-accretion regime). However, pseudo-evolution can account for most of the mass change during the late stages of halo evolution when the actual accretion of matter slows down or stops altogether. In a $\Lambda$-dominated universe, accretion stops entirely for some halos \citep{prada_06_outerregions, diemand_07_haloevolution, cuesta_08_infall}, and eventually for all halos \citep{busha_05}. The fraction of halos in the slow-accretion regime thus depends on mass, redshift, and cosmology. At $z\lesssim 1$, pseudo-evolution is dominant for galaxy-sized halos but dominates even for up to a quarter of cluster-sized halos, namely those with the earliest formation epochs \citep[D13,][]{wu_13_rhapsody1}. In our previous study, we showed that the evolution of the concentration-mass relation of halos during the late stages of their evolution can almost entirely be attributed to pseudo-evolution of the halo radius (D13). This example illustrates that it is important to consider possible deviations from the predictions of the self-similar collapse model, which does not explicitly account for the effects of the mass definition.

In this paper, we focus on clusters of galaxies and quantify the effects of pseudo-evolution on cluster scaling relations. As an example, we consider the scaling relation between halo mass and velocity dispersion, both defined within the halo radius. The $M_{\Delta} - \sigma(<R_{\Delta})$ relation can be measured in $N$-body simulations, and it has been shown that cluster-sized cold dark matter (CDM) halos exhibit a tight power-law relation between $M$ and $\sigma$ consistent with the self-similar expectation \citep[][hereafter E08]{evrard_08},
\begin{equation}
\label{eq:msigmarel}
\sigma(<R_{\Delta})\propto\sqrt{\frac{G \, M(<R_{\Delta})}{R_{\Delta}}}\propto \left[ \rhot(z)^{1/2} M_{\Delta}(z) \right]^{1/3} .
\end{equation}
The slope of the \msr of cluster-sized halos in simulations has been shown to be close to the expected value of $1/3$ (E08). We note that the scaling with $\rhot$ in Equation (\ref{eq:msigmarel}) takes into account the changing relation between $M_{\Delta}$ and $R_{\Delta}$ due to the spherical overdensity mass definition in Equation (\ref{eq:sodef}), but not the different ways in which pseudo-evolution could affect $M_{\Delta}$, $R_{\Delta}$, and $\sigma(<R_{\Delta})$. Thus, we could expect that halos that grow mostly through pseudo-evolution might deviate from the scaling relations of halos that physically accrete matter at a high rate. We choose the \msr as a test case because it can be studied with dissipationless simulations, allowing us to use larger simulation volumes and larger, statistical samples of cluster halos. However, the effects of pseudo-evolution we study are generic, and we discuss their implications for other cluster scaling relations. 

This paper is organized as follows. In Section \ref{sec:results}, we compare the evolution of the \msr of halo samples that have undergone different amounts of pseudo-evolution, and we interpret the results in Section \ref{sec:interpretation}. In Section \ref{sec:otherrel}, we discuss the impact of pseudo-evolution on other scaling relations. We summarize our findings in Section \ref{sec:conclusion}. Throughout the paper, we denote the generic radius from the center of a cluster halo as $r$, whereas we reserve capital $R$ for specific radii used to define halo mass. Quantities that are integrated or averaged over a spherical volume are indicated as $<r$, such as $M(<r)$ or $\sigsr$, whereas $\sigr$ refers to the velocity dispersion {\it at} a particular radius $r$. We denote the mean matter density of the universe $\rho_{\rm m}$, and the critical density $\rho_{\rm c}$. Mass definitions referring to $\rho_{\rm m}$ or $\rho_{\rm c}$ are understood to have a fixed overdensity $\Delta$, and are denoted $M_{\Delta \rm m} = M(<R_{\Delta \rm m})$, such as $\mtom$, or $M_{\Delta \rm c} = M(<R_{\Delta \rm c})$, such as $\mtoc$. The labels $\mvir$ and $\rvir$ are reserved for a varying overdensity $\Delta(z)$ with respect to the matter density, where $\Delta_{\rm vir}(z=0) \approx 358$ and $\Delta_{\rm vir}(z>2) \approx 180$ for the cosmology assumed in this paper \citep[e.g.,][]{bryan_98_virial}. Last, the more general threshold density $\rhot$ denotes the product $\Delta(z) \rho_{\rm ref}(z)$ for any given mass definition, and it is understood to be a function of redshift.


\begin{deluxetable*}{lccccccccl}
\tablecaption{Best-fit Power Laws to the $\mtoc-\sigma_{\rm 200c}$ Relation
\label{table:fits}}
\tablewidth{0pt}
\tablehead{
\colhead{Sample} &
\colhead{$z$} &
\colhead{$N$} &
\colhead{$N/N_{\rm tot}$} &
\colhead{$M_{\rm Pivot}$} &
\colhead{$\sigma_{\rm Pivot}$} &
\colhead{Offset} &
\colhead{$\sigma(M=14.2)$} &
\colhead{Slope} &
\colhead{Intrinsic Scatter}
}
\startdata
All halos                    & $0$    & $13412$ & $100\%$ & $14.2275$ & $2.7704$ & $0.0 \pm 0.0001$ & $2.7610 \pm 0.0001$ & $0.3436 \pm 0.0007$ & $0.0165 \pm 0.0001$ \\
$0.00 < \fpe \leq 0.25$ (FA) & $0$    & $1567$   & $11.7\%$ & $14.2868$ & $2.7964$ & $0.0 \pm 0.0006$ & $2.7671 \pm 0.0005$ & $0.3369 \pm 0.0022$ & $0.0216 \pm 0.0004$ \\
$0.25 < \fpe \leq 0.50$        & $0$    & $5235$   & $39.0\%$ & $14.2388$ & $2.7762$ & $0.0 \pm 0.0003$ & $2.7630 \pm 0.0003$ & $0.3409 \pm 0.0012$ & $0.0182 \pm 0.0002$ \\
$0.50 < \fpe \leq 0.75$        & $0$    & $2934$   & $21.9\%$ & $14.2233$ & $2.7690$ & $0.0 \pm 0.0003$ & $2.7611 \pm 0.0003$ & $0.3417 \pm 0.0013$ & $0.0137 \pm 0.0002$ \\
$0.75 < \fpe \leq 1.00$ (PE) & $0$    & $1363$   & $10.2\%$ & $14.1970$ & $2.7573$ & $0.0 \pm 0.0003$ & $2.7583 \pm 0.0003$ & $0.3430 \pm 0.0018$ & $0.0111 \pm 0.0002$ \\
$1.00 < \fpe < \infty$          & $0$    & $1826$   & $13.6\%$ & $14.1859$ & $2.7501$ & $0.0 \pm 0.0002$ & $2.7550 \pm 0.0002$ & $0.3438 \pm 0.0015$ & $0.0103 \pm 0.0002$ \\
$-\infty < \fpe \leq 0$         & $0$    & $487$      & $3.6\%$ & $14.1732$ & $2.7421$ & $0.0 \pm 0.0005$ & $2.7512 \pm 0.0005$ & $0.3424 \pm 0.0036$ & $0.0116 \pm 0.0004$ \\
All halos                      & $0.5$ & $13374$  & $99.7\%$ & $13.9922$ & $2.7282$ & $0.0 \pm 0.0002$ & $2.7627 \pm 0.0002$ & $0.3434 \pm 0.0006$ & $0.0184 \pm 0.0001$ 
\enddata
\tablecomments{The fits are performed in the $\log_{10} M - \log_{10} \sigma$ plane and all values listed above refer to $\log_{10}$ of the corresponding values. The pivot point of the fit was set to the logarithmic average $M$ and $\sigma$ of the fitted sample, and the listed offset corresponds to the difference from this pivot in $\sigma$. For easier comparison of the normalizations, $\log_{10} \sigma$ at $\log_{10} M = 14.2$ is also listed. The uncertainty on this normalization depends on the uncertainties in the offset and slope, as well as the difference between the pivot mass and $14.2$. The normalization of the best-fit to the $z = 0.5$ sample is $\log_{10} \sigma = 2.7996$, while the value listed above is re-scaled to the $z = 0$ value expected from Equation (\ref{eq:msigmarel}), i.e. $\mtoc \rightarrow \mtoc \times E(z)$, and thus $\sigma \rightarrow \sigma \times \alpha E(z).$, with $\alpha = 0.3434$. The normalization, slope and scatter of the $z = 0$ samples are visualized in Figure \ref{fig:msigmarels}.}
\end{deluxetable*}

\section{Pseudo-evolution and the \ms Relation}
\label{sec:results}

In this section, we investigate the evolution of simulated halos in the \ms plane, comparing halo samples that have experienced different amounts of physical mass accretion. 

\subsection{Numerical Simulation and Methods}
\label{sec:results:sim}

We use a dissipationless simulation of a $1 \gpccubed$ cubic volume in the \LCDM model. The large box size ensures that the volume contains tens of thousands of cluster-sized halos. For consistency with the analysis of D13, we chose the same cosmological parameters as the Bolshoi simulation \citep{klypin_11_bolshoi}, namely a flat $\Lambda$CDM model and $\Omega_\rmm=1-\Omega_\Lambda=0.27$, $\Omega_\rmb=0.0469$, $h=H_0/(100\kmsmpc)=0.7$, $\sigma_8=0.82$ and $n_\rms=0.95$. These parameters are compatible with measurements from WMAP7 \citep{jarosik_11_wmap7}, a combination of WMAP5, Baryon Acoustic Oscillations and Type Ia supernovae \citep{komatsu_etal11}, X-Ray cluster studies \citep{vikhlinin_09_clusters}, and observations of the clustering of galaxies and galaxy-galaxy/cluster weak lensing \citep[see, e.g.,][]{tinker_12_galcosmo, cacciato_13}. The same cosmology was used for all analytical calculations in this paper. The initial conditions for the simulation were generated at redshift $z=49$ using a second-order Lagrangian perturbation theory code \citep[2LPTic,][]{crocce_06_2lptic}. The simulation was run using the publicly available code Gadget2 \citep{springel_05_gadget2}, and followed $1024^3$ dark matter particles, corresponding to a mass resolution of $7 \times 10^{10} \msunh$. The spline force softening was set to $1/30$ of the inter-particle separation, $33 \kpch$. 

We used the phase-space halo finder Rockstar \citep{behroozi_13_rockstar} to extract all isolated halos and subhalos from the 100 snapshots of the simulation, and applied the merger tree code of \citet{behroozi_13_trees} to these halo catalogs. Whenever we refer to the progenitor of a halo, we mean the halo along its most massive progenitor branch at each redshift. We focus on a sample of isolated cluster-sized halos with $\mvir > 5 \times 10^{13} \msunh$ at $z = 0$, consisting of $\approx 57,000$ halos, though we impose further mass cuts for some of the experiments below. For 99.9\% of the halos a $z=1$ progenitor could be identified, and for 98.6\% of them a $z=2$ progenitor was found.  In the following analysis, only progenitors with $M > 10^{13} \msunh$ (referring to whichever mass definition used) were considered, corresponding to about 140 particles in the smallest halos. Once this lower limit was reached, earlier progenitors were also removed from the analysis. For all halos in the sample, we extract spherically averaged density profiles in 80 logarithmically spaced bins between $0.05 \rvir$ and $10 \rvir$. 

For the fits to the \msr in Section \ref{sec:results:msr}, we also follow E08 in excluding satellite halos, i.e. halos whose $\rtoc$ overlaps with a larger halo. This requirement excludes more halos than just subhalos that are required to have their {\it centers} lie inside $\rtoc$ of a larger halo. We find the satellite fraction to be a few percent depending on the mass range, somewhat lower than E08. The difference is presumably attributable to the different halo finders and cosmologies used.

Since we are interested in subtle differences between the \msr of different halo samples, we need to measure the slope and offset of the power-law fits (i.e., linear fits in log-log space) robustly. For the \ms data at hand, a simple least-squares fit is not appropriate because of the presence of intrinsic scatter \citep{hogg_10_fitting}. Instead, we perform a Markov-Chain Monte Carlo analysis with 30,000 samples for each fit, sampling the likelihood of the offset, slope, and intrinsic scatter of the linear fit in the $\log M-\log\sigma$ plane.

\begin{figure*}
\centering
\includegraphics[trim = 11mm 20mm 8mm 3mm, clip, scale=0.66]{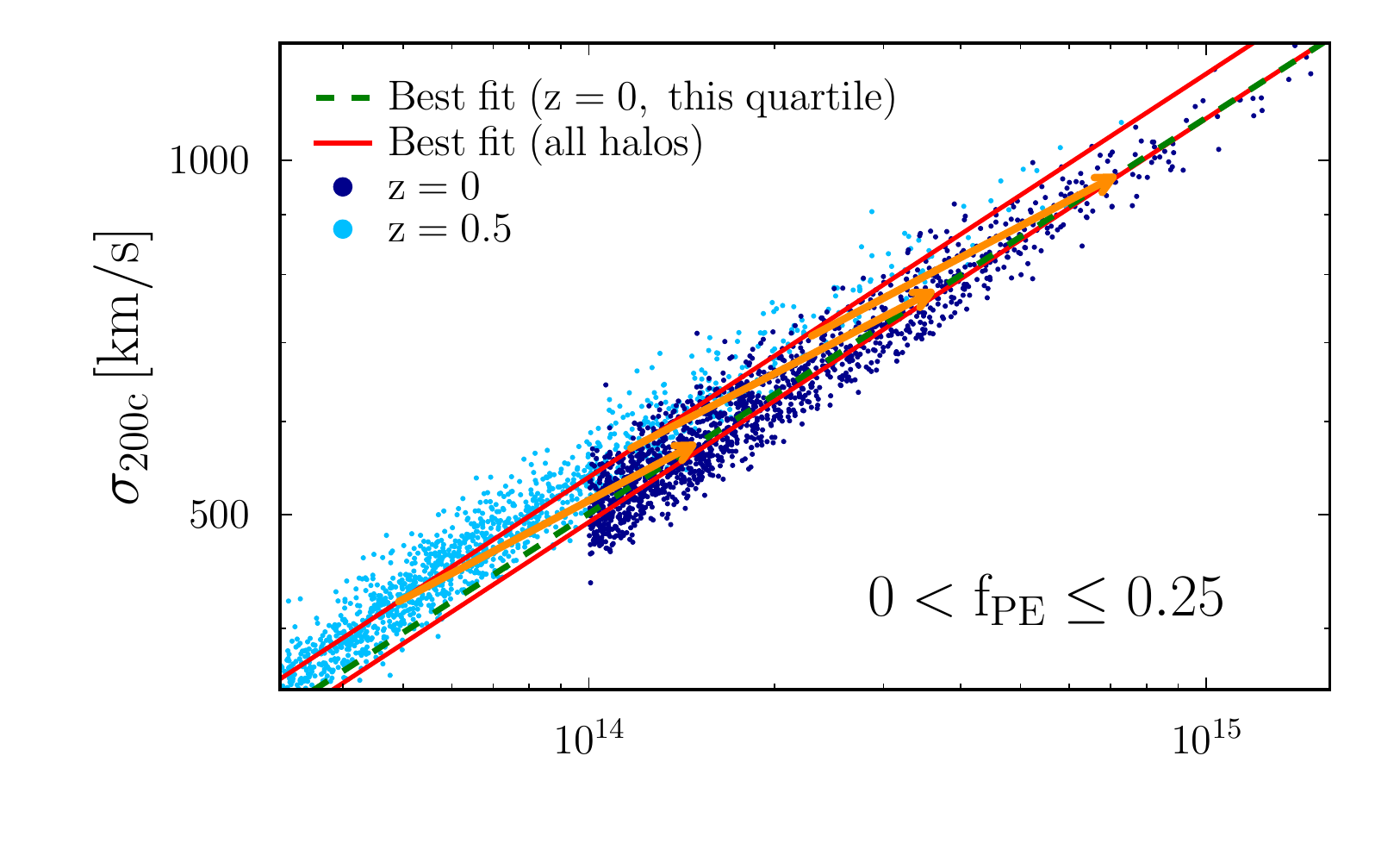}
\includegraphics[trim = 32mm 20mm 7mm 3mm, clip, scale=0.66]{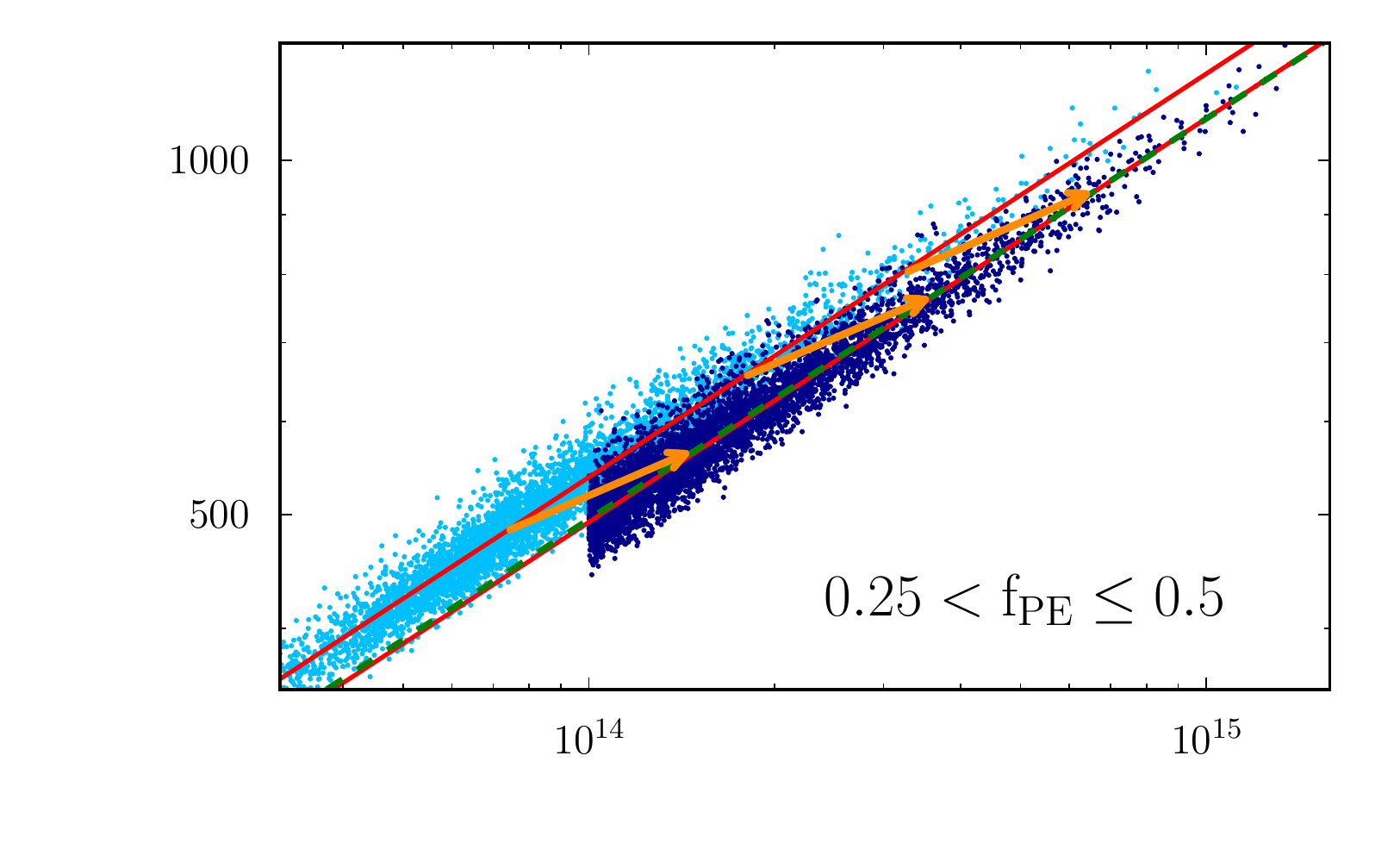}
\includegraphics[trim = 11mm 3mm 8mm 4mm, clip, scale=0.66]{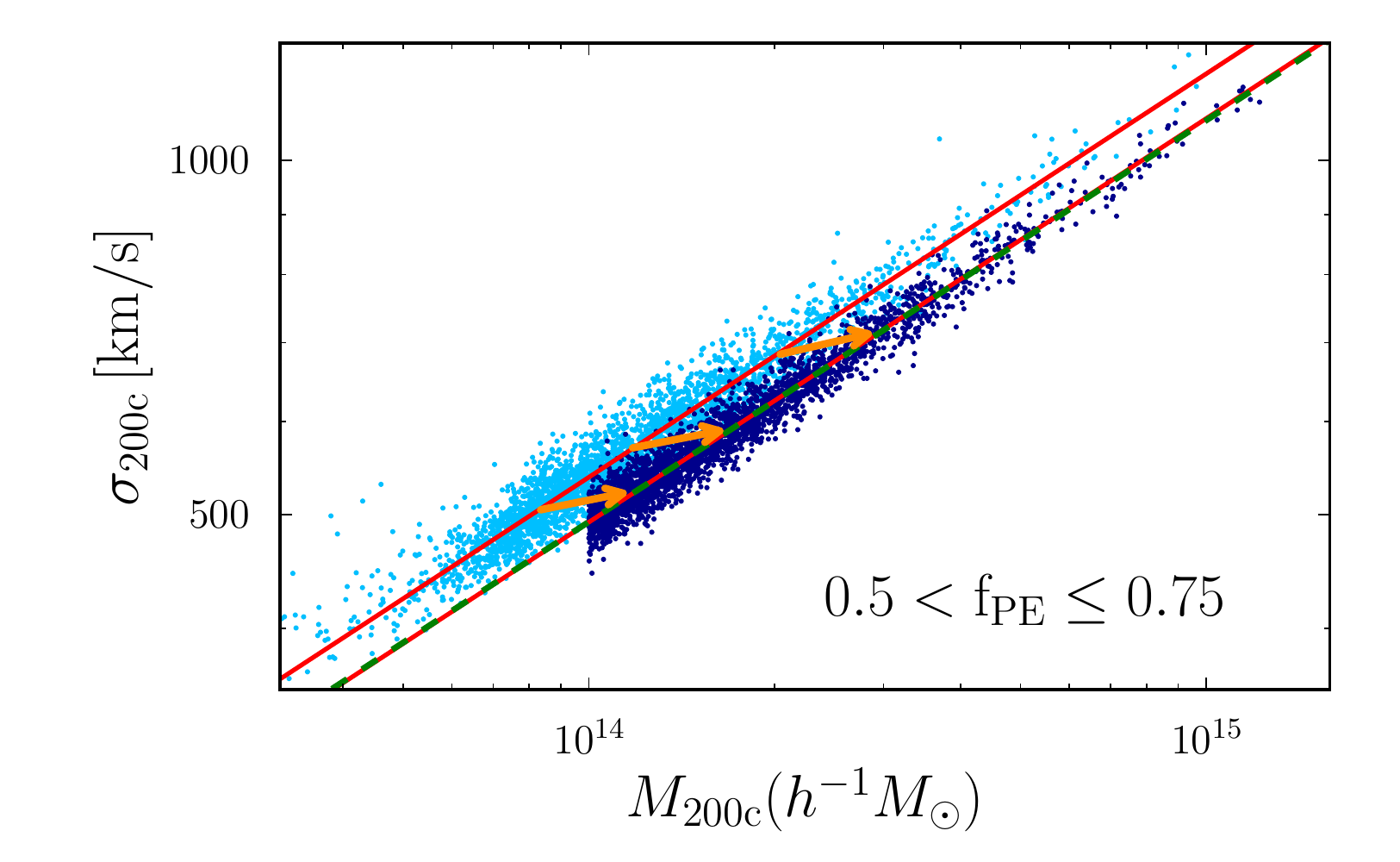}
\includegraphics[trim = 32mm 3mm 7mm 4mm, clip, scale=0.66]{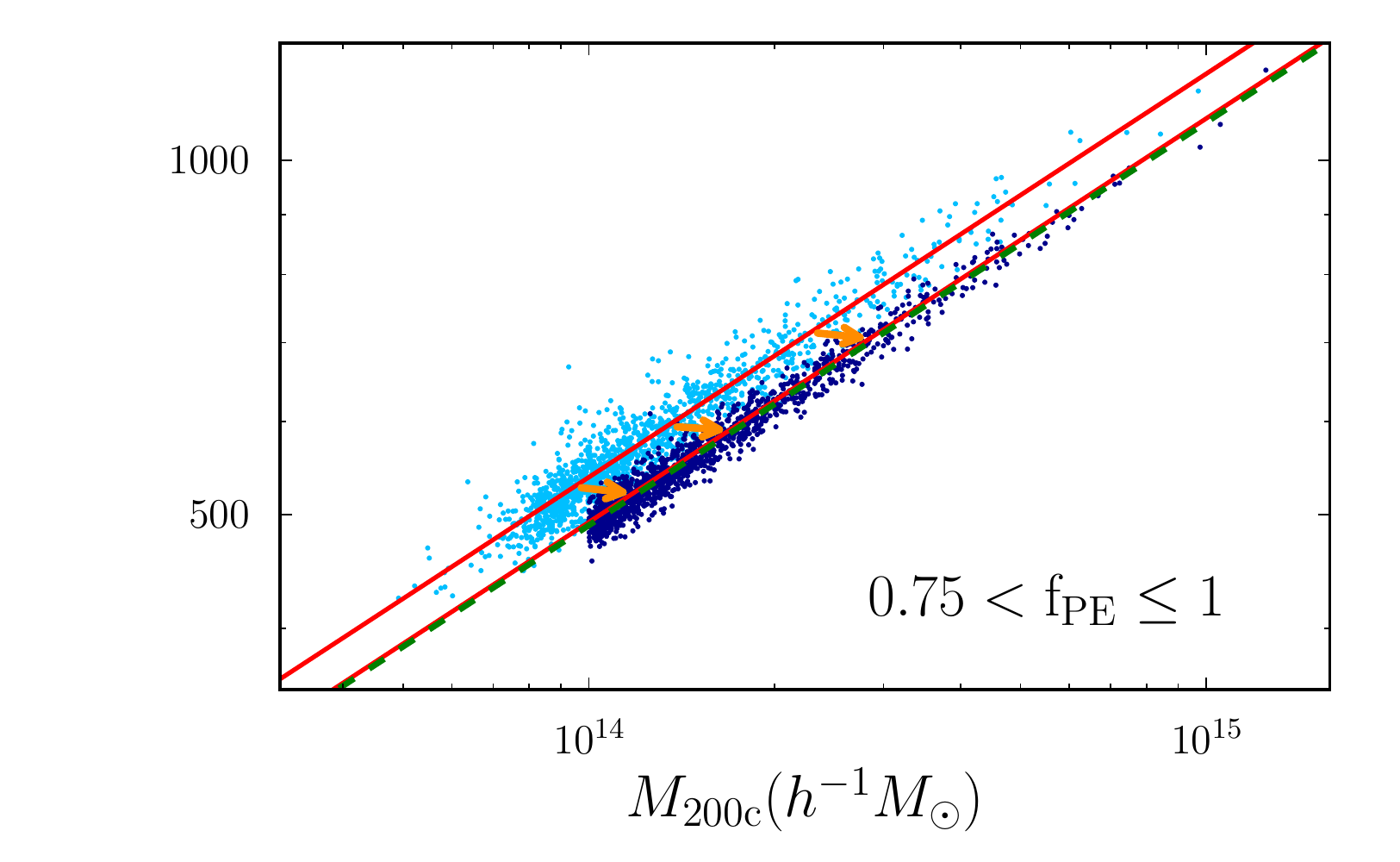}
\caption{Evolution of cluster-sized halos ($\mtoc >10^{14}\msunh$ at $z = 0$) on the $M-\sigma$ plane. The halos are split into four samples according to their fraction of pseudo-evolution between $z=0.5$ and $0$, $\fpe$, with the lowest $\fpe$ in the top left panel, and the largest $\fpe$ in the bottom right panel. The dark blue and light blue dots indicate the positions of halos at $z=0$ and $0.5$, respectively. The orange arrows indicate the evolution of the logarithmic average of halos in three mass bins. The lines show the best-fit power-law relations to the entire sample at $z=0$ and $z=0.5$ (red), as well as a best-fit to the halos in the respective sub-sample of each panel (green dashed). The physically accreting halos (top panels) increase both their $M$ and $\sigma$ with time, while those halos that grow mostly through pseudo-evolution (bottom panels) shift approximately horizontally.}
\label{fig:msigevo}
\end{figure*}

\subsection{The Pseudo-evolution of Halo Mass}
\label{sec:results:pe}

The spherical overdensity mass definition given in Equation (\ref{eq:sodef}) is based either on the critical or mean matter density of the universe, both of which decrease with time, albeit at different rates. As a result, the threshold density $\rhot$ decreases, even for the virial mass definition where the evolution of $\Delta_{\rm vir}$ and $\rhoc$ partially cancel. Thus, the virial radius and mass of a halo increase, even if the halo does not evolve and its density profile remains constant. The amount of this pseudo-evolution of mass can be computed for a given density profile by numerically solving Equation (\ref{eq:sodef}) with a varying density threshold $\rhot(z)$. D13 found pseudo-evolution to account, on average, for a growth of about a factor of two in $\mtom$ since $z = 1$, albeit with large scatter. 

In the simple case of a static density profile, the amount of pseudo-evolution can be determined exactly if the density profile is known at any epoch. We sometimes refer to this case as ``pure pseudo-evolution''. In the more realistic case where a halo grows through a mixture of physical accretion and pseudo-evolution, the exact amount of pseudo-evolution can only be estimated, unless the density profile is known at all redshifts. In this paper, we use the minimum estimator of pseudo-evolution defined in D13, which corresponds to the mass between the virial radii at some initial redshift, $R(z_{\rm i})$, and at a final redshift, $R(z_{\rm f})$, which is already in place at $z_{\rm i}$. This material is bound to contribute to the halo's pseudo-evolution as the virial radius moves out (see D13 for details on rare exceptions). As the density between the two virial radii increases after $z_{\rm i}$, the actual amount of pseudo-evolution is larger. Thus, this estimator represents the minimum pseudo-evolution we deduce from the density profiles at $z_{\rm i}$ and $z_{\rm f}$. We define the fraction of mass growth due to pseudo-evolution, $\fpe$, as the minimum pseudo-evolved mass divided by the actual difference in spherical overdensity mass between the main progenitor at $z_{\rm i}$ and its descendant at $z_{\rm f}$ (see D13 for an exact mathematical definition).

Most galaxy-sized halos are predominantly pseudo-evolving after $z \approx 1$ \citep[][]{cuesta_08_infall}. Moreover, the mass growth is dominated by pseudo-evolution for a non-negligible fraction of cluster-sized halos at $z<1$ \citep[D13]{wu_13_rhapsody1}. In this paper, we use $z_{\rm i} = 0.5$ and $z_{\rm f} = 0$ as the interval during which we estimate the amount of pseudo-evolution. This interval corresponds to about $5\, \rm Gyr$, or a couple of dynamical times of clusters. 

\subsection{The Evolution of the $M-\sigma$ Relation}
\label{sec:results:msr}

\begin{figure}
\centering
\includegraphics[trim = 2mm 0mm 2mm 2mm, clip, scale=0.7]{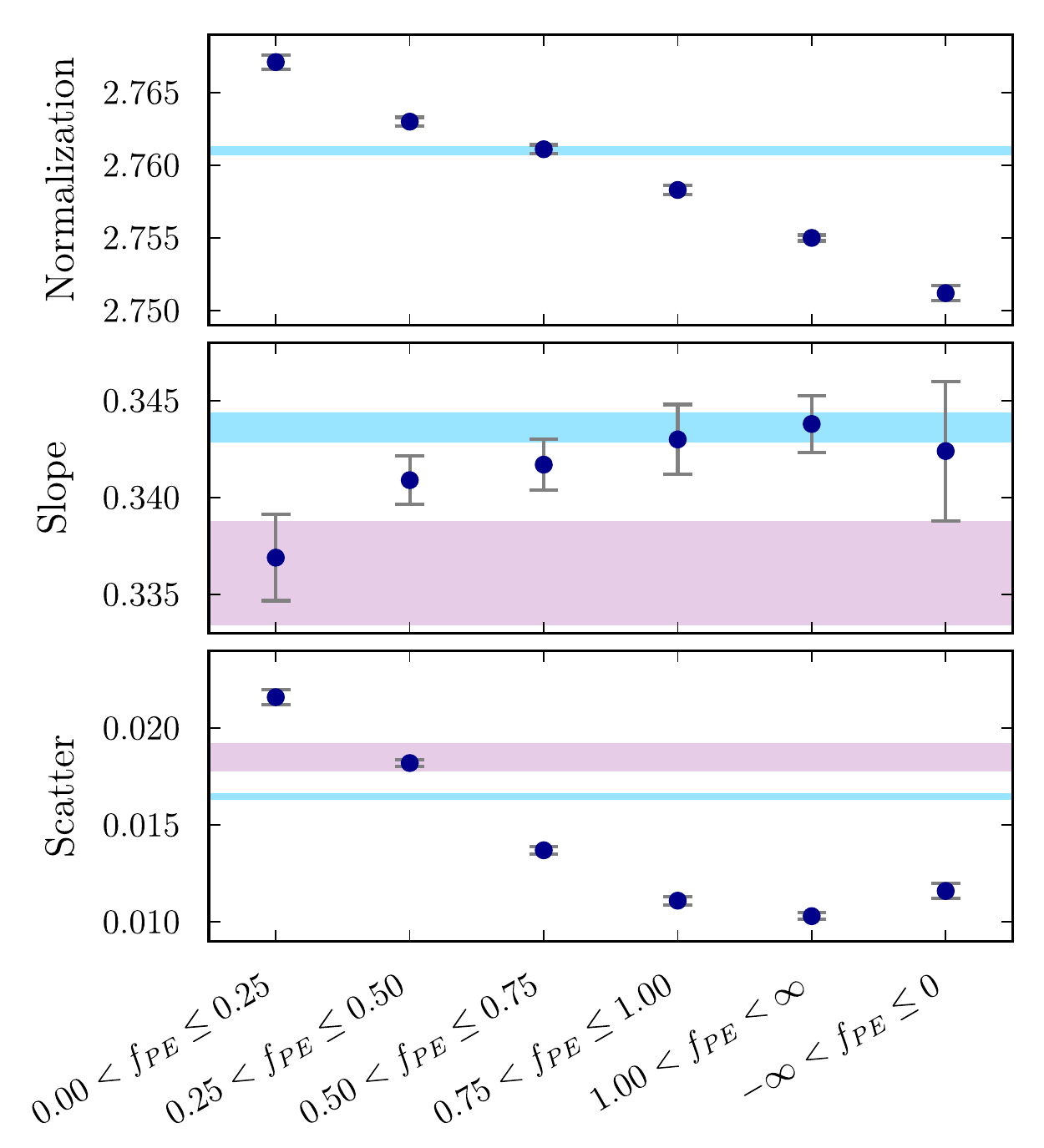}
\caption{ Normalization, slope and scatter of the \msr for the different halo samples listed in Table \ref{table:fits}. The light-blue bands correspond to the $1\sigma$ regions of the values for the fit to all halos at $z = 0$, while the dark blue points show the values for the respective sub-samples. The purple bands show the best-fit values of E08 for comparison.}
\label{fig:msigmarels}
\end{figure}

We first examine the relation between $\mtoc$ and $\sigma_{\rm 200c}$, as E08 found that the \msr is particularly tight for this mass definition. We define $\sigma$ as the averaged, one-dimensional, velocity dispersion inside a radius $r$ (E08),
\begin{equation}
\label{eq:sigmadef}
\sigsr \equiv \left[ \frac{1}{3 N_{\rm p}(r)} \sum_{i=1}^{N_{\rm p}(r)} \sum_{j=1}^{3} \left| v_{\rm i,j}-\bar{v}_{\rm j} \right|^2\right]^{\frac{1}{2}}
\end{equation}
where $N_{\rm p}(<r)$ is the number of particles within $r$, $v_{\rm i,j}$ the $j$th component of the velocity of particle $i$, and $\bar{v}_{\rm j}$ the average $j$th component of all particles inside $r$. 

We consider all isolated halos with $\mtoc >10^{14}\msunh$ at $z = 0$, a sample of $\approx 13,400$ halos. We split those halos into six samples with different $\fpe$, the fraction of their mass evolution since $z = 0.5$ estimated to be due to pseudo-evolution. Table \ref{table:fits} lists the fraction of halos in each sample; for the halos with $\fpe > 1$, the density profile decreases at certain radii. For a small fraction of halos, $\mtoc$ decreases with time, and thus $\fpe < 0$. For the rest of the paper, we refer to halos with $0.75 < \fpe < 1$ as pseudo-evolving (PE), and those with $0 < \fpe < 0.25$ as fast accreting (FA).

Figure \ref{fig:msigevo} shows the \ms evolution of halos in the four samples spanning the range $0 < \fpe < 1$. The lines show the best-fit power-law relation to the entire sample at $z=0$ and $z=0.5$ (solid red lines), as well as to the halos in the respective $\fpe$ sample at $z = 0$ (dashed green lines). The orange arrows indicate the average evolution of halos in three mass bins at $z = 0$. Table \ref{table:fits} lists the parameters of the best-fit power laws for the respective halo samples, some of which are visualized in Figure \ref{fig:msigmarels}. 

In general, the \msr of all halos in our simulation at $z = 0$ roughly agrees with the results of E08, and matches the redshift scaling predicted by Equation (\ref{eq:msigmarel}). There are, however, some subtle disagreements (Figure \ref{fig:msigmarels}). First, our best-fit slope $\alpha \equiv d {\ln} \sigma / d {\ln} M=0.3436 \pm 0.0007$ is steeper than the slope $\alpha = 0.3361 \pm 0.0026$ found by E08. It is interesting to note that the E08 slope is consistent with the expected value of $1/3$, whereas our slope differs from this value by $14\sigma$. One might suspect that the sharp mass cut-off at $10^{14} \msunh$ introduces an artificial steepening, but the progenitor halos at $z = 0.5$ are not subject to such a cut, and their \msr exhibits the same slope as at $z = 0$, within the statistical uncertainties (Figure \ref{fig:msigevo}). Furthermore, the mass resolution of our simulation is sufficiently high to measure the slope of the \msr without bias (Figure 6 of E08; our mass resolution corresponds to 14,000 particles in a $10^{15} \msunh$ halo). In Section \ref{sec:interpretation:profiles}, we present a physical explanation for the steepening of the slope, as well as for the trends in normalization and slope with $\fpe$. A further, minor difference appears in the intrinsic scatter about the best-fit relation that we find to be $0.0165$ dex, while E08 reported a somewhat larger scatter of $0.0185$ dex.

It is remarkable that although halos with different amounts of pseudo-evolution evolve very differently on the \ms plane, the \ms relations of the PE and FA samples evolve nearly identically. The slopes of the relation for the individual samples agree with the slope for the entire sample to better than $2\%$, with the largest disagreement in the FA sample (about $3\sigma$). The slope of the best-fit relation to the PE sample agrees with the overall sample within statistical errors. The difference in the normalization between the FA and PE sets is roughly $2\%$ as well. Thus, we conclude that pseudo-evolution does not have a significant impact on the \msr for $\mtoc$. The main difference between the samples is that the scatter around the \msr is noticeably smaller for the PE sample, likely because the PE halos are the most relaxed sub-population.


\section{The Origin of the \ms Relation and its Evolution}
\label{sec:interpretation}

The results presented in the previous section show that one \msr describes all halos across a wide spectrum of accretion states, even the extreme cases of halos that are mostly growing through pseudo-evolution, and those that are accreting mass at high rates. In this section, we explore the origin of this remarkable similarity.

\subsection{Preliminary Considerations}
\label{sec:interpretation:prelim}

Figure \ref{fig:msigevo} clearly shows that both FA and PE halos in simulations (with masses defined within $\rtoc$) follow a power-law \msr,
\begin{equation}
\label{eq:msigmapowerlaw}
\sigma = \sigma_M^* S(z) \left( \frac{M}{M_*} \right)^{\alpha} \,.
\end{equation}
The slope of the best-fit power law is close to $\alpha = 1/3$, and the evolution of the relation is consistent with $S(z) = \rhot^{1/6}$, the scaling expected from the simple estimate described in Section \ref{sec:intro}. We first consider the origin of this relation for FA halos.

It is well known that the density and mass profiles of CDM halos are universal as a function of radius scaled by  the scale radius $\rs$,
\begin{equation}
M(<r) = M_{\Delta} \frac{\mu(x)}{\mu(c)}
\end{equation}
where $x = r/\rs$, $\rs$ is the radius where the density profile has a logarithmic slope of $-2$, and $c = R_{\Delta} /\rs$ is a dimensionless concentration parameter. In addition to the universal mass profile, CDM halos exhibit a power-law pseudo-phase space density profile within the radius of the first shell crossing,
\begin{equation}
\label{eq:qprofile}
Q(r) \equiv \frac{\rho(r)}{\sigma^3(r)} = Q_* x^{\beta} \,,
\end{equation}
with $\beta \approx 1.9$ \citep{taylor_01, ascasibar_04, rasia_04, ludlow_11}. The existence of universal density and $Q$ profiles implies that $\sigsr$ also follows a universal profile,
\begin{equation}
\label{eq:sigsc1}
\sigsr = \sigma_\ast s(x)
\end{equation}
where $s(x)$ is another dimensionless function of the re-scaled radius. By dimensional considerations\footnote{There are four physical quantities in the problem,  $\sigma$, $G$, $M$, and $R$. Their units are composed of three physical dimensions, length, mass, and time. Hence, by the $\Pi$-theorem, there is only one dimensionless variable that can be formed, $\sigma/\sqrt{GM/R}$, which implies a relation $\sigma=s\sqrt{GM/R}$, where $s$ is a dimensionless function of dimensionless parameters.}
\begin{equation}
\label{eq:sigsc2}
\sigma_\ast = \sqrt{\frac{G \, M(< r)}{r}}
\end{equation}
where $r$ is a control radius of the problem. Given that $\rs$ appears to be the control radius in the density structure of halos, $r$ can be readily identified with $\rs$. Thus, 
\begin{eqnarray}
\label{eq:sigsc3}
\sigma(<r) &=& \sqrt{\frac{G \, M(< r_s)}{r_s}}\, s(x) \nonumber\\
&=&\sqrt{\frac{G \, \mdelta}{\rdelta}}\, s(x)\left[\frac{\mu(c)}{c\,\mu(1)}\right]^{-1/2}
\end{eqnarray}
which allows us to express $\sigsr$ in terms of the scaling of Equation (\ref{eq:msigmarel}),
\begin{equation}
\label{eq:sigsc4}
\sigma(<\rdelta) \propto \rhot^{1/6}(z) \mdelta^{1/3} \times s(c) \, \left[\frac{\mu(c)}{c\,\mu(1)}\right]^{-1/2} \,.
\end{equation}
For FA halos, $\cvir = \rvir / \rs$ is approximately constant, $\cvir\approx 4$ \citep{zhao_03_concentration, zhao_09_mah}, and thus $\sigma(<\rvir) \propto \rhot^{1/6}(z) \mvir^{1/3}$ according to Equation (\ref{eq:sigsc4}). Note that this result does not depend on the actual shape of $\mu(x)$ and $s(x)$, but only on their universality. 

Often, however, the \msr is defined within some radius $\rdelta \neq \rvir$, e.g. $\rfoc$. In such cases, $c$ generally evolves with redshift (particularly at $z\lesssim 1$), and we would expect some evolution in the normalization of $\sigma$ in addition to $\rhot^{1/6}(z)$. Nevertheless, Figure~\ref{fig:msigevo} shows no such evolution, which means that the $s(x)/[\mu(x)/x]^{1/2}$ term in Equation (\ref{eq:sigsc4}) must be approximately constant for different $x\sim\rdelta$. We assume that the $\sigsr$ and $M(<r)$ profiles can be approximated by power laws around $\rdelta$,
\begin{equation}
\label{eq:powerlawdefs}
\mu(x) = \tilde\mu x^{\gamma} \,; \ \ \ \  s(x) = \tilde s x^{\xi} \, .
\end{equation}
The condition that $s(x)/[\mu(x)/x]^{1/2} \approx$ constant is satisfied if 
\begin{equation}
\xi-\frac{\gamma-1}{2}\approx 0 \,; \,\,\,\, \tilde{\mu}/\tilde s^{1/2} \approx {\rm constant} \,.
\label{eq:slope_rel}
\end{equation}
Thus, we expect $\sigma \propto \rhot^{1/6}(z)M^{1/3}_{\Delta}$ for any $\rdelta$ as long as the $M$ and $\sigma$ profiles satisfy the conditions expressed in Equation (\ref{eq:slope_rel}). We note that these conditions are valid for both PE halos and for halos that evolve through a combination of physical accretion and pseudo-evolution. Their $\sigma_\ast$ stops increasing, and their $\rdelta$ increases as a result of pseudo-evolution; but as long as the $M$ and $\sigma$ profiles obey the relations in Equation (\ref{eq:slope_rel}), the PE halos move along the same \msr as the FA halos.

\subsection{The $M$ and $\sigma$ Profiles in Simulations}
\label{sec:interpretation:profiles}

\begin{figure}
\centering
\includegraphics[trim = 0mm 15mm 0mm 1mm, clip, scale=0.62]{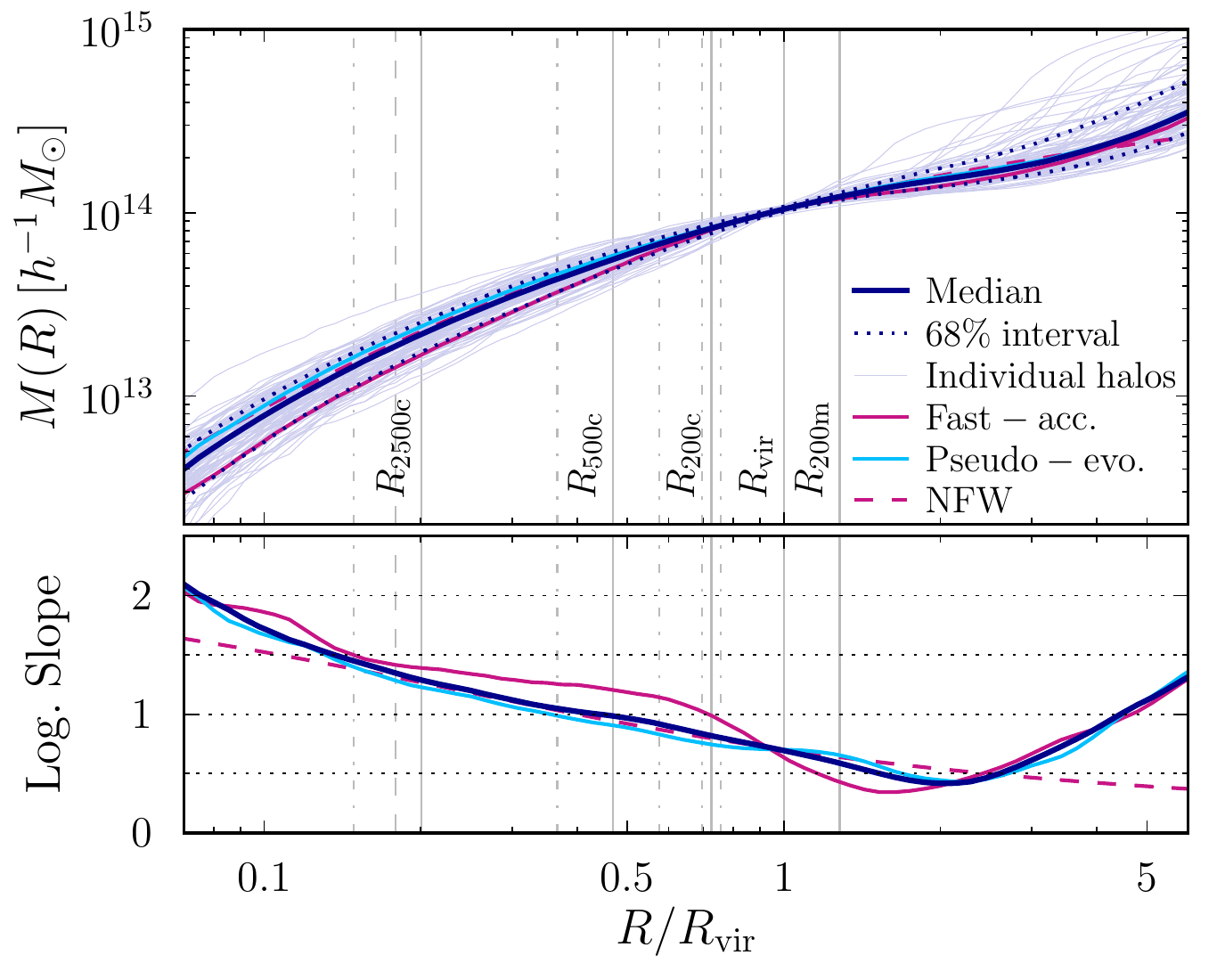}
\includegraphics[trim = 0mm 15mm 0mm 2mm, clip, scale=0.62]{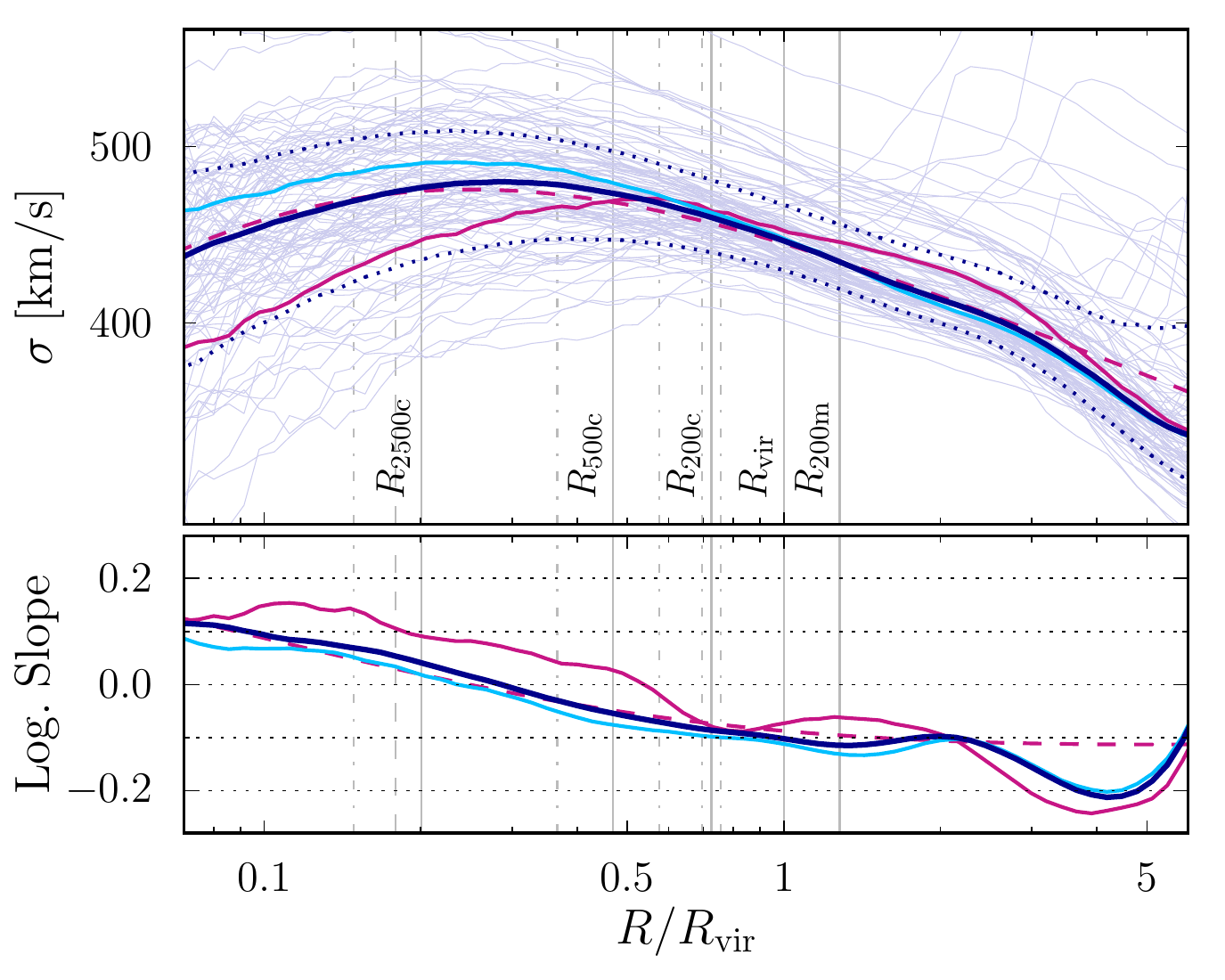}
\includegraphics[trim = 0mm 3mm 0mm 2mm, clip, scale=0.62]{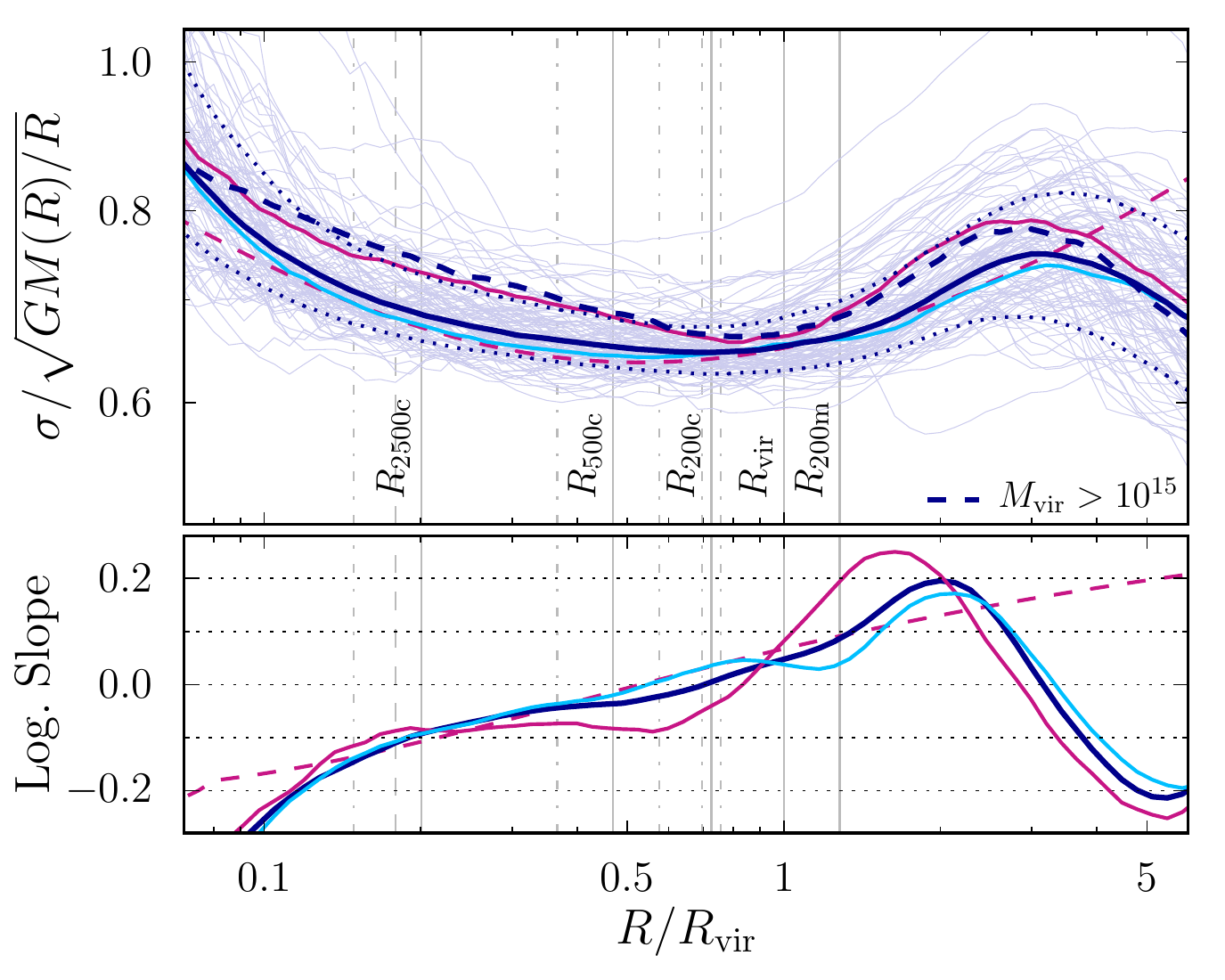}
\caption{Radial profiles and logarithmic slopes of $M(<r)$ (top panels), $\sigsr$, (center panels), and $\sp \equiv \sgmr$ (bottom panels) for isolated halos in a narrow mass bin with $10^{14} \msunh < \mvir < 1.1 \times 10^{14} \msunh$ (about 3,000 halos). The profiles of about 100 randomly selected individual halos are plotted in gray. The median and 68\% contours are shown in dark blue. The sample shown contains halos with all fractions of pseudo-evolution. For comparison, the slope of the mostly pseudo-evolving (PE) sample ($0.75 < \fpe \leq 1$) is plotted in light-blue lines, and the slope of the fast-accreting (FA) sample ($0 < \fpe \leq 0.25$) in purple lines. The slope of an NFW density profile with a concentration typical for this halo mass is plotted with a dashed purple line. The virial radii for five mass definitions, $\rtfc$, $\rfoc$, $\rtoc$, $\rvir$ and $\rtom$, are plotted as vertical gray lines, with solid lines indicating the radii at $z = 0$ and dot-dashed lines at $z = 0.5$. The dashed gray line close to $\rtfc$ marks the scale radius, $r_{\rm s}$. In the bottom panel, the dashed dark blue line shows the median $\sp$ for halos with $\mvir > 10^{15} \msunh$. As most of those halos are fast-accreting, their $\sp$ profile resembles that of the FA sample at $10^{14} \msunh$.}
\label{fig:profiles}
\end{figure}

We now investigate whether the $M$ and $\sigma$ profiles of CDM halos in simulations obey Equation (\ref{eq:slope_rel}). Figure \ref{fig:profiles} shows the mass and velocity dispersion profiles of halos in a narrow mass bin with $10^{14} \msunh < \mvir < 1.1 \times 10^{14} \msunh$, a sample of about 3,000 halos. For simplicity, we focus on the profiles at $z = 0$. The profiles are plotted for $r>0.07 \, \rvir$ which corresponds to about twice the softening length of our simulation. We have examined the profiles of massive halos, $\mvir > 10^{15} \msunh$, which are resolved with ten times more particles and have larger $\rvir$. We found very similar profile shapes in $M$ and $\sigma$, which indicates that 1) the shapes are not strongly dependent on halo mass, and 2) the inner structure of the profiles for the $10^{14} \msunh$ halos does not suffer from significant resolution effects in the radial range shown.

Figure \ref{fig:profiles} shows the median profiles for all halos, as well as the FA and PE sub-samples, along with the logarithmic slopes of the median profiles as a function of radius. The slopes were computed using a fourth-order Savitzky-Golay smoothing algorithm over the 15 nearest bins \citep{savitzky_64} to smooth out random fluctuations, but without affecting the actual values of the slope. 

In order to investigate deviations from the scaling of Equation (\ref{eq:msigmarel}), the bottom panel of Figure \ref{fig:profiles} shows the profile of the scaled velocity dispersion $\sp \equiv \sgmr$. For the cluster halos examined here, the median $\sp$ is approximately constant from $\rtfc$ to $\rtom$ -- i.e., the range of radii usually used to define $M$ and $\sigma$. Thus, the first condition of Equation (\ref{eq:slope_rel}) is satisfied over these radii. For example, at $\rtoc$, $\gamma = 0.82$ and $\xi = -0.09$, and thus $\xi - (\gamma-1)/2 \approx 0$. The second condition of Equation (\ref{eq:msigmarel}) demands that the normalization of the $\sp$ profiles of halos with different masses and accretion rates is roughly constant. This condition is more or less satisfied, but we note two deviations. First, the FA halos (purple line in the bottom panel of Figure \ref{fig:profiles}) have a slightly higher median $\sp$ profile than the median and the PE halos. Second, higher-mass halos ($\mvir > 10^{15} \msunh$, dashed dark blue line) follow the profile of the lower-mass FA halos almost exactly, which is not surprising since most high-mass halos are in the FA regime.

These slight deviations from the conditions of Equation (\ref{eq:slope_rel}) explain the deviations of the \msr from a perfectly self-similar scaling that we observed in Section \ref{sec:results:msr}. First, the higher median $\sp$ profiles of higher-mass halos mean that the slope of the \msr of the entire sample is steeper than $1/3$. The difference in the normalization appears to be larger at small radii such as $\rtfc$ (Figure \ref{fig:profiles}), and we measure the slope of the \msr for $\mtfc$ to be steeper, about $0.36$. Second, the higher normalization of $\sp$ of the FA compared to the PE halos explains why the \ms relations for the FA samples have systematically higher normalizations than those of the PE samples (Figure \ref{fig:msigmarels}). Finally, we note that the slopes measured {\it within each sample} also show a trend: For the FA sample, the slope is consistent with $1/3$, but increases significantly toward the PE sample. This trend confirms that FA halos adhere to conditions of Equation (\ref{eq:slope_rel}) most accurately, as expected from the discussion in Section \ref{sec:interpretation:prelim}. The steeper slope within the PE sample could, for example, arise if the lowering of the $\sp$ normalization happens gradually, and the higher-mass halos in a given PE sample have, on average, resided in the PE regime for less time than have their lower-mass counterparts. 

Despite the small deviations discussed earlier, the profiles are close to a constant $\sp$ with radius and mass. This property of the profiles is the main reason why the FA and PE halos evolve similarly, and why the cluster population exhibits a tight scaling relation of the form $\sigma \propto \rhot^{1/6}(z)M^{1/3}$ for all common definitions of $\mdelta$.  There are, however, small deviations from a constant $\sp$ at small and large radii such as $\rtfc$ and $\rtom$. We examine the implications of these deviations for the \msr in Section \ref{sec:interpretation:massdefinition}.

\subsection{The Origin of the Profiles}
\label{sec:interpretation:origin}

What is the origin of the remarkable constancy of $\sp = \sgmr$ with radius? We observe significant systematic differences in the shapes of the $M(<r)$ and $\sigsr$ profiles of the FA and PE halos shown in Figure~\ref{fig:profiles}, and note that these differences are real, and not due to noise in the measurement. Nevertheless, $\sp(r)$ is similar for both types of halos. 

First, we note that the $\sigsr$ profile is well described by a simple prediction based on the Jeans equation. Specifically, we assume that halos follow the Navarro-Frenk-White (NFW) density profile,
\begin{equation}
\rho_{\rm NFW} = \frac{\rho_0}{x(1+x)^2} \,,
\end{equation}
and thus
\begin{equation}
M_{\rm NFW} = 4 \pi \rs^3 \rho_0 f(x)
\end{equation}
where $f(x) = \ln(1 + x) - x / (1 + x)$. We assign the NFW profile the same virial radius as the median profile, and the mean concentration, $c_{\rm vir}=\rvir / r_{\rm s} = 5.6$, corresponding to the median mass, $\mvir = 1.05 \times 10^{14} \msunh$ of our halo sample. This mean concentration was derived from the concentration-mass relation of \citet{zhao_09_mah}, which was calibrated with cosmological N-body simulations. The top panel of Figure \ref{fig:profiles} demonstrates that the NFW profile, shown with a dashed purple line, is a good approximation to the median mass profile of simulated halos at the radii of interest. 

To compute the $\sigma$ profile corresponding to an NFW density profile, we assume spherical symmetry,  Jeans equilibrium,\footnote{The term ``Jeans equilibrium'' means that, at a given radius, the static Jeans equation holds \citep{binney_08_galdyn}. The system may be in Jeans equilibrium locally at some radii, while out of equilibrium at other radii. Virial equilibrium, in contrast, applies to a system as a whole and requires a vanishing second derivative of the global moment of inertia tensor.} and that the orbits of dark matter particles in cluster halos are only mildly anisotropic across most radii within the virial radius \citep{eke_98, colin_00, cuesta_08_infall, lemze_12}. We thus find \citep[see, e.g.,][]{more_09_sk1}
\begin{equation}
\sigma^2(<r) = \frac{c_{\rm vir} V_{\rm vir}^2}{f(c_{\rm vir}) f(x)} \int_0^{x} x'^2 dx' \int_{x'}^{\infty} \frac{f(x'') dx''}{x''^3(1+x'')^2} \,.
\end{equation}
This predicted profile is a good approximation both to the median $\sigma(<r)$ profile (center panel of Figure \ref{fig:profiles}), and to the corresponding profile of the PE sample, even though most halos in this mass range are undergoing physical mass accretion. This equilibrium estimate offers a possible explanation why the FA and PE halos have similar $\sp$ profiles, even though their $M$ and $\sigma$ profiles are noticeably different. This view is further supported by the fact that halos maintain power-law $Q(r)$ profiles during different stages in their evolution. We can conclude, therefore, that halos maintain a structure close to equilibrium even during stages of rapid mass growth.

We now generalize the NFW prediction to any power-law density profile, $\rho=\rho_0 x^{\alpha}$ where $\alpha\ne -1$. The mass enclosed within the scaled radius, $x$, is 
\begin{equation}
M(<x) = \frac{4 \pi \rho_0 \rs^3}{(\alpha+3)} x^{\alpha+3} = M_0 x^{\alpha+3}\,.
\end{equation}
The velocity dispersion profile at a given radius $r$, using the spherically symmetric Jeans equation, is given by
\begin{equation}
\sigma^2(x) = -\frac{GM_0}{\rs} \frac{x^{\alpha+2}}{(2\alpha+2)}\,.
\end{equation}
Again assuming isotropic orbits, we obtain the mass-weighted velocity dispersion averaged within the scaled radius,
\begin{eqnarray}
\sigma^2(<x) &=& -\frac{GM_0}{\rs}\frac{1}{(2\alpha+2)} \frac{\int x'^{\alpha+2} x'^\alpha x'^2 dx' }{\int x'^\alpha  x'^2 dx'} \\
&=& -\frac{G M(<x)}{\rs x}\frac{(\alpha+3)}{(2\alpha+2)(2\alpha+5)} \,.
\end{eqnarray}
The aforementioned calculations show that $\sigma(<r) \approx {\rm const}\times \sqrt{GM(<r)/r}$ for both NFW profiles, and profiles that roughly follow a power law around the relevant radii. We conclude that a relation of the form $\sigma \propto \rhot^{1/6}(z) M^{1/3}$ can be generically expected for CDM halos, as long as they are in approximate Jeans equilibrium.

The results presented in this section illustrate the origin of tight scaling relations, such as the \msr: they reflect equilibrium relations between cluster profiles. The scaling relations hold at any radius where such equilibrium relations apply.

\subsection{The Impact of Mass Definition}
\label{sec:interpretation:massdefinition}

\begin{figure}
\centering
\includegraphics[trim = 2mm 19mm 4mm 0mm, clip, scale=0.52]{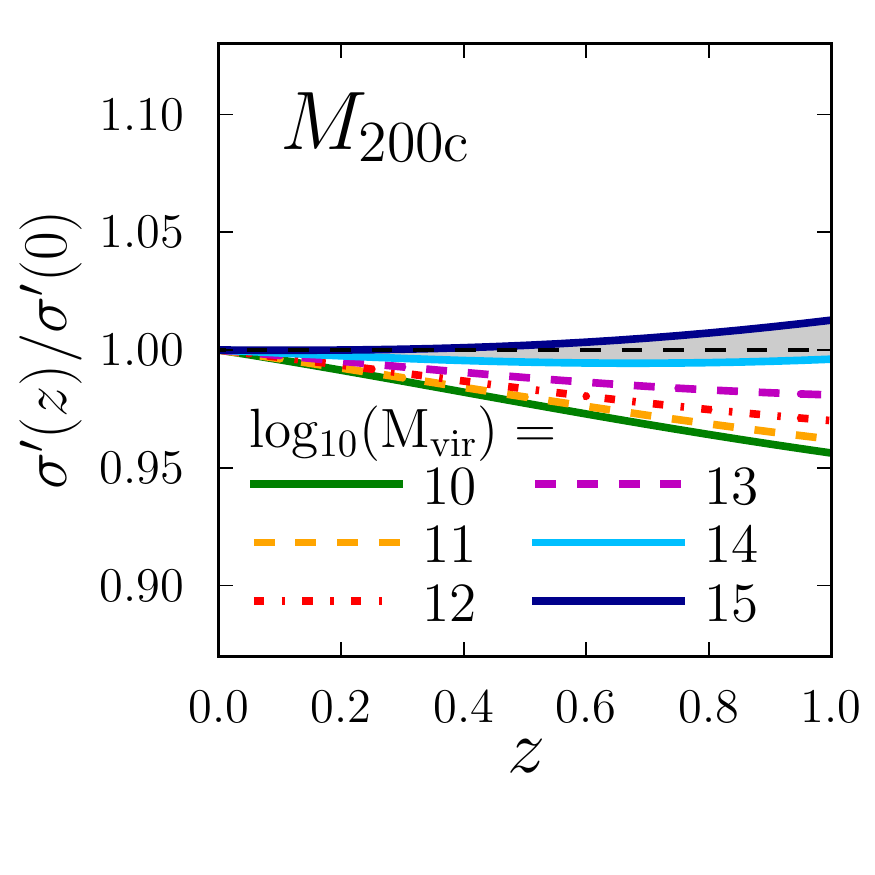}
\includegraphics[trim = 19mm 19mm 1mm 0mm, clip, scale=0.52]{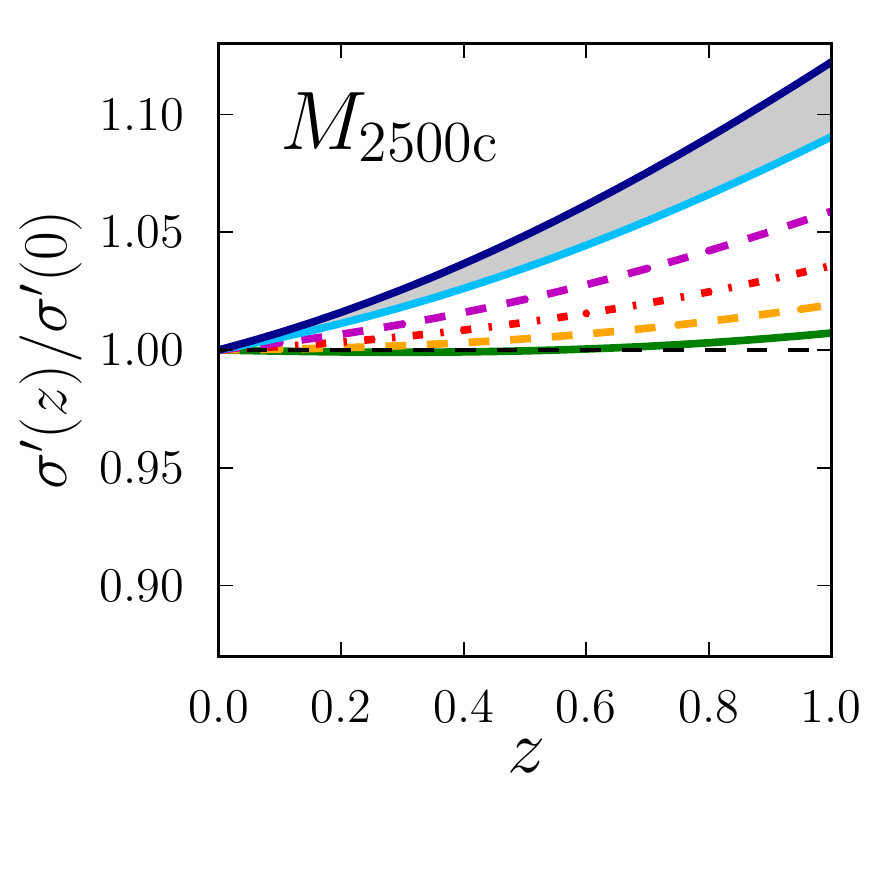}
\includegraphics[trim = 2mm 9mm 4mm 4mm, clip, scale=0.52]{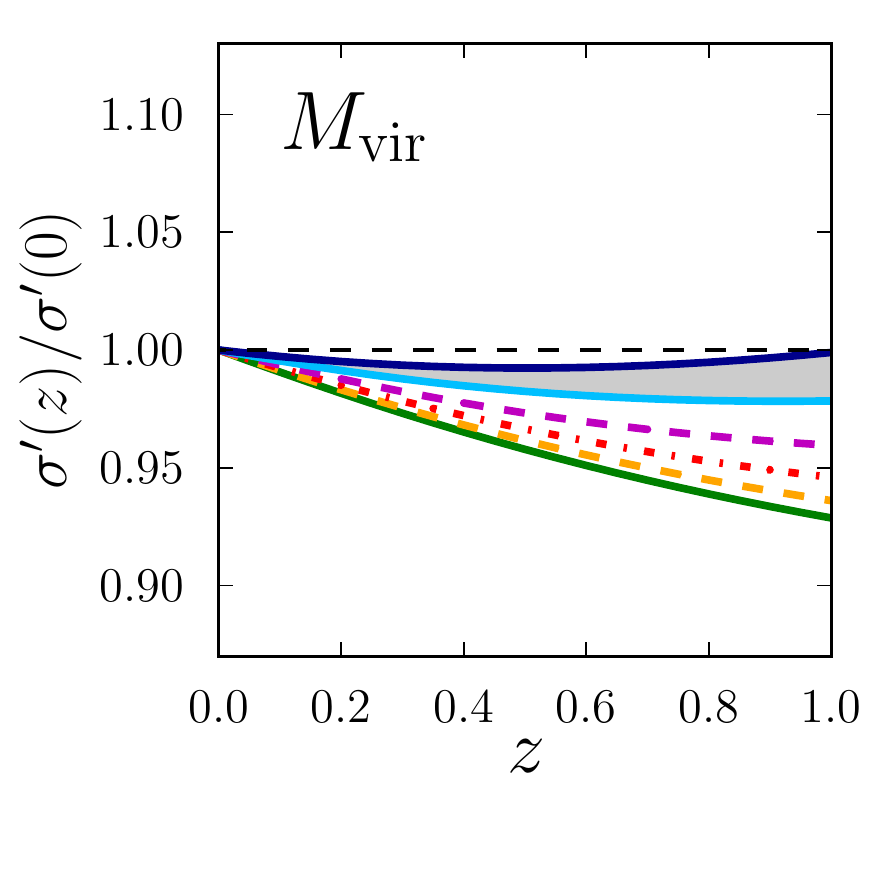}
\includegraphics[trim = 19mm 9mm 1mm 4mm, clip, scale=0.52]{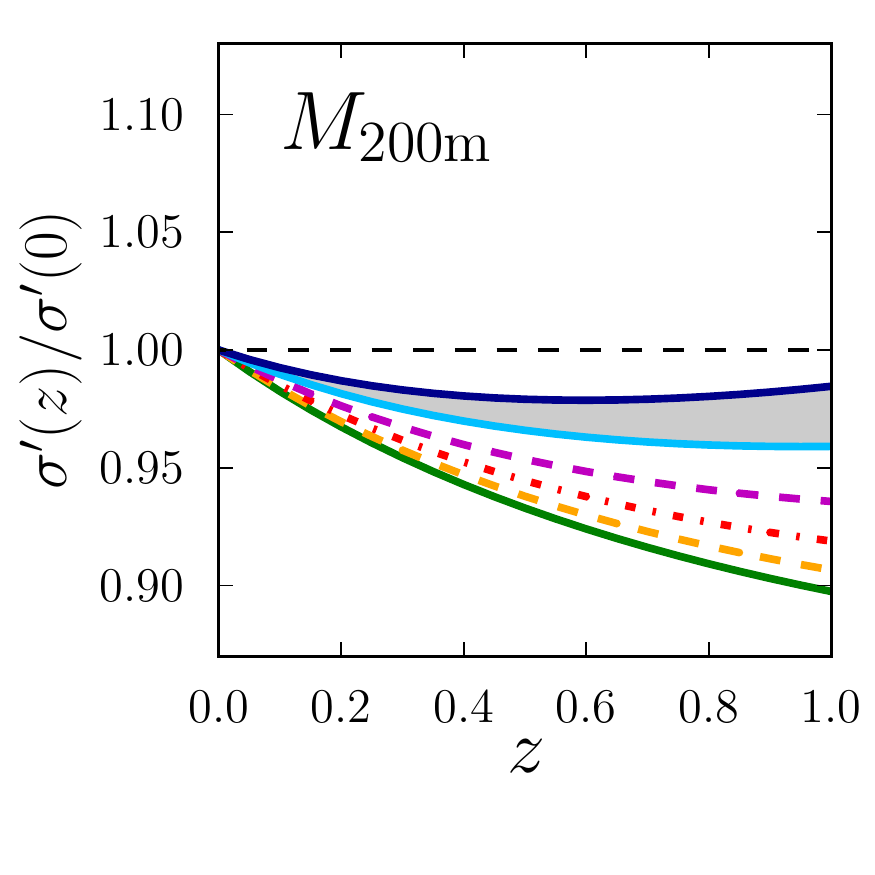}
\caption{Pseudo-evolution of the re-scaled velocity dispersion, $\sp = \sgmr$, for various mass definitions. The evolution was computed assuming that halos follow static NFW density profiles and the concentration-mass relation given by the model of \citet{zhao_09_mah}. The lines refer to different masses at $z=0$, all measured as $\mvir$. The gray shaded area highlights the mass range of the majority of cluster halos, between $10^{14}$ and $10^{15} \msunh$.}
\label{fig:analytical}
\end{figure}

\begin{figure}
\centering
\includegraphics[trim = 11mm 2mm 5mm 3mm, clip, scale=0.58]{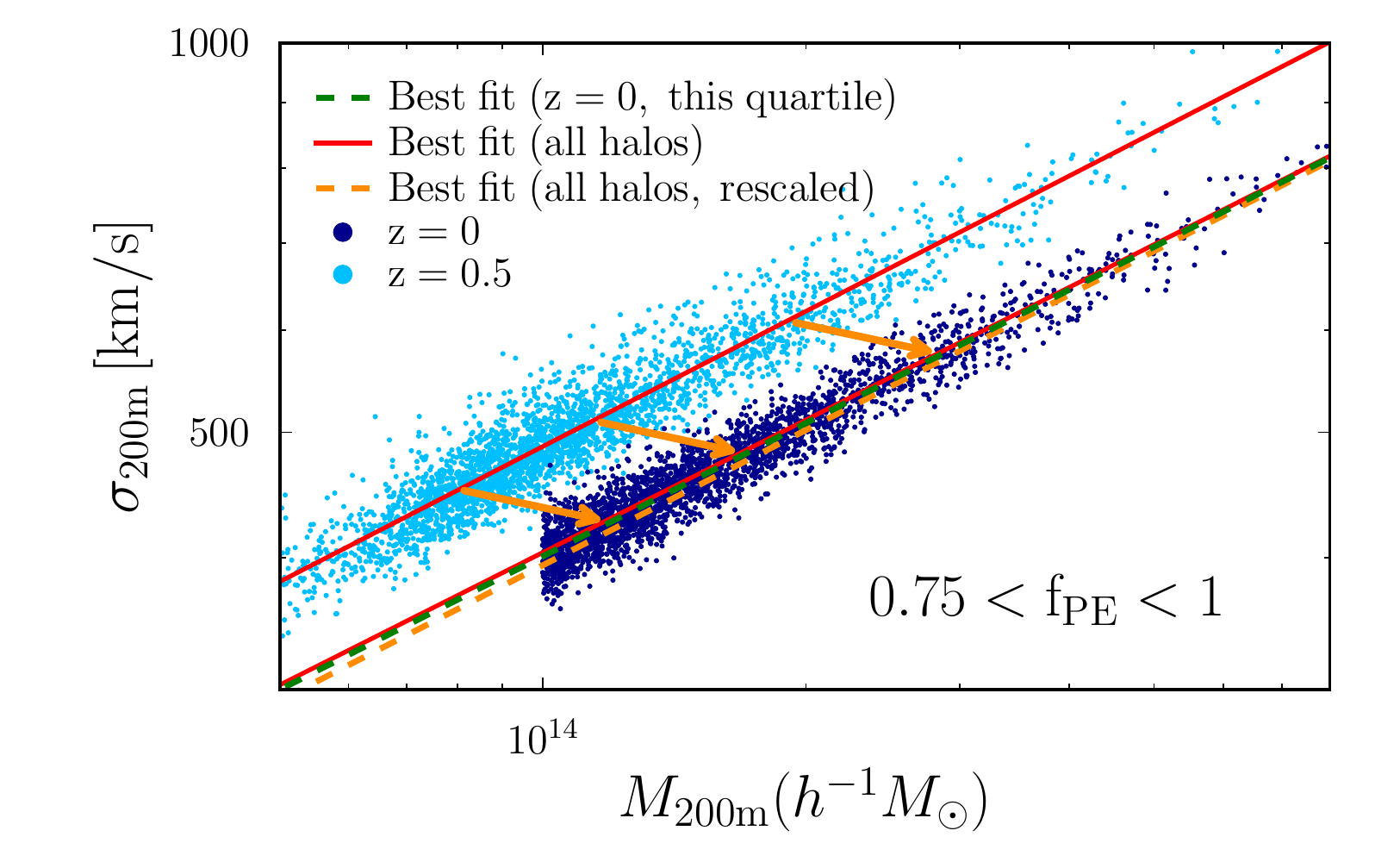}
\includegraphics[trim = 10mm 3mm 5mm 3mm, clip, scale=0.58]{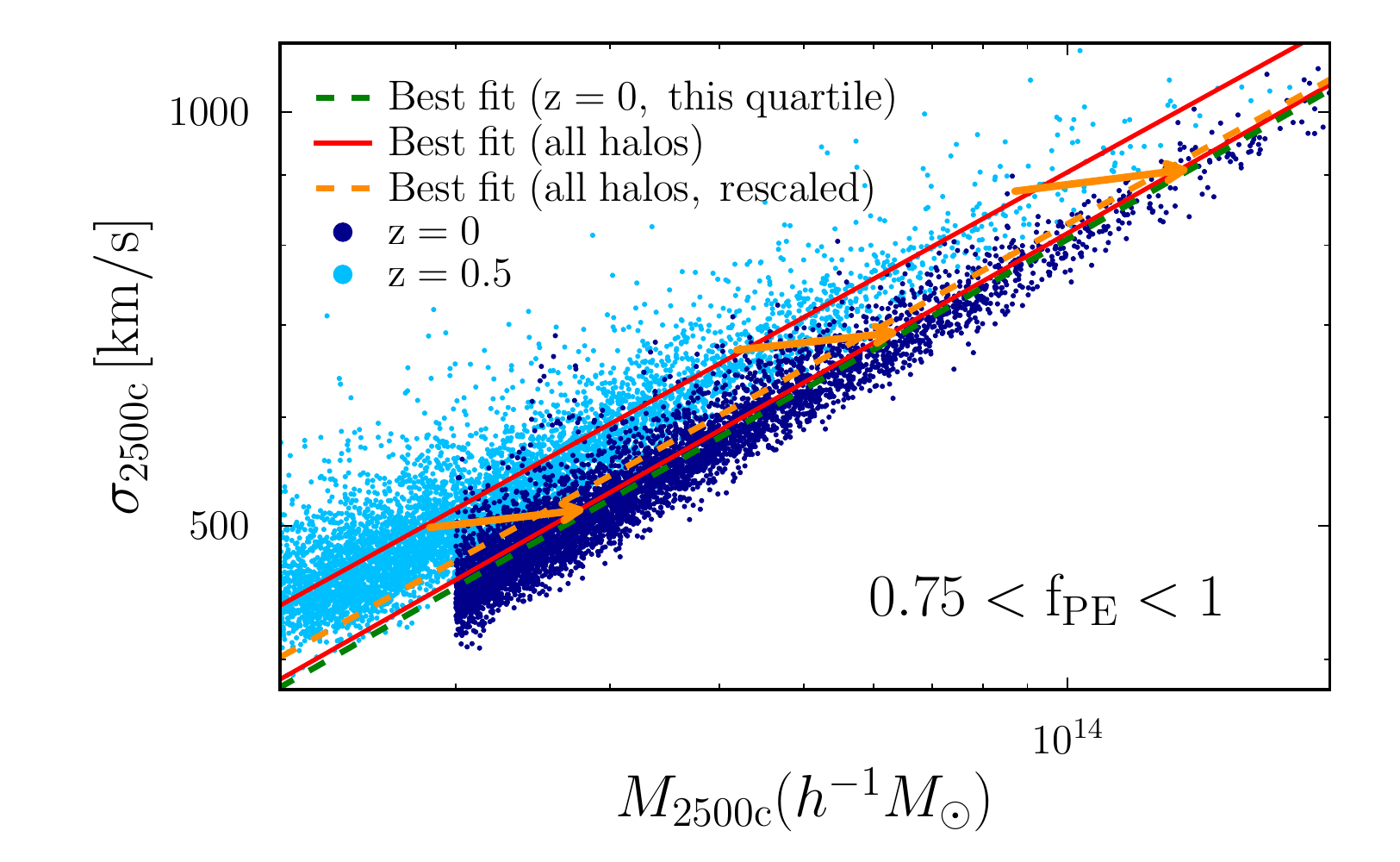}
\caption{Same as Figure \ref{fig:msigevo}, but showing only the PE sample of halos, and the $\mtom$ (top panel) and $\mtfc$ (bottom panel) mass definitions. As for the $\mtoc$ mass definition, the fits to the PE sample (green dashed lines) are not noticeably offset from the overall sample (red lines). However, the redshift evolution of the \msr deviates from the expected scaling (dashed orange line). See Section \ref{sec:interpretation:massdefinition} for a detailed discussion.}
\label{fig:msigevo2}
\end{figure}

We now return to the small deviations from a constant $\sp$ observed in Figure \ref{fig:profiles}, which can cause a difference in the evolution of the \msr for FA and PE halos, depending on the chosen mass definition. To evaluate these deviations, we use the NFW model and the $\sp$ profiles calculated using Jeans equilibrium and the assumption of isotropic orbits, which proved to be a good approximation to the median $\sp$ profile (Section \ref{sec:interpretation:origin}).
We assign concentrations appropriate for a halo of a given mass using the model of \citet{zhao_09_mah}. Figure \ref{fig:analytical} shows the evolution of $\sp$ from $z=0$ to $z=1$ for four different mass definitions, normalized to $z=0$. Each color corresponds to a certain $\mvir$ (regardless of the mass definitions used in the respective panels). The resulting physical density and $\sigma$ profiles are assumed to be constant over time. The gray shaded region highlights the mass range $10^{14} \msunh < \mvir < 10^{15} \msunh$ in which most cluster halos reside. An \msr that is perfectly independent of radius, and thus of pseudo-evolution, corresponds to the dashed horizontal line indicating $\sp =$ const. 

The results shown in Figure \ref{fig:analytical} are a reflection of the slope of the $\sp$ profile shown in Figure \ref{fig:profiles}, and the positions of the respective virial radii. The top left panel demonstrates that $\sp$ is particularly constant for large halo masses and an $\mtoc$ mass definition. As expected, the deviations from a perfect scaling are small, $\Delta \sp \lesssim 10\%$ at $z = 1$. The figure illustrates that there are two main factors that influence the impact of pseudo-evolution on $\sp$, namely (1) the chosen definition of the virial radius, and (2) the concentration of the NFW profile, and thus the halo mass. For mass definitions with a low threshold density ($\mtom$, $\mvir$), the  $\sp$ profile around the virial radius has a positive slope and $\sp$ increases between $z = 1$ and $z = 0$ for pseudo-evolving halos. Conversely, for mass definitions with high threshold densities ($\mfoc$, $\mtfc$), the slope is negative and $\sp$ decreases with time. We note that, for most mass definitions $\sp$ is generally less constant for smaller mass halos because they have larger concentrations and their virial radii shift further along the $\sp$ profile. For most mass definitions this shift leads to larger deviations from a flat slope, but $\rtfc$ moves into a more favorable part of the profile as the concentration increases. Thus, we conclude that it is somewhat fortuitous that the \msr for cluster halos is close to the simple scaling of Equation (\ref{eq:msigmarel}), as lower-mass halos should evolve to deviate from the self-similar prediction for the normalization by $\approx 10\%$, depending on the mass definition.

Figure \ref{fig:msigevo2} shows the evolution of the PE halos on the \ms plane for the $\mtom$ and $\mtfc$ mass definitions. As $\mtfc$ is significantly smaller than $\mtoc$, the cut-off mass for the $\mtfc$ sample was lowered to $\mtfc \geq 2 \times 10^{13} \msunh$. As expected from Figure \ref{fig:analytical}, halos evolve in different directions in $\sigma$ in the top and bottom panels. The fits to the PE sample (green dashed lines) are almost indistinguishable from the fits to all halos (solid red lines), as for the $\mtoc$ mass definition in Figure \ref{fig:msigevo}.  However, the \msr deviates somewhat from the expected redshift scaling; the dashed orange lines show the fit to the $z = 0.5$ sample re-scaled to $z = 0$ according to Equation (\ref{eq:msigmarel}). For the $\mtom$ mass definition, this scaling underpredicts the measured $\sigma$ at fixed mass. We expect this deviation, since the $\sp$ profile increases with radius around $\rtom$ and $\sigma$ at $z = 0$ will thus be larger than we would expect from a constant $\sp$ scaling. Conversely, the $\sp$ profile decreases with radius around $\rtfc$, meaning that Equation (\ref{eq:msigmarel}) overpredicts the measured $\sigma$ for this mass definition. These findings underscore that, for certain mass definitions, pseudo-evolution can have an impact on the \msr of the {\it entire population}, not only the PE halos.

We conclude that the \msr exhibits an evolution close to that expected from the scaling relation of Equation (\ref{eq:msigmarel}) only for definitions of the virial radius that lie in the flat range of the $\sp$ profile. Other definitions of masses and radii will result in small deviations from such a scaling.


\section{Implications for other cluster scaling relations}
\label{sec:otherrel}

\begin{figure}
\centering
\includegraphics[trim = 10mm 5mm 2mm 2mm, clip, scale=0.68]{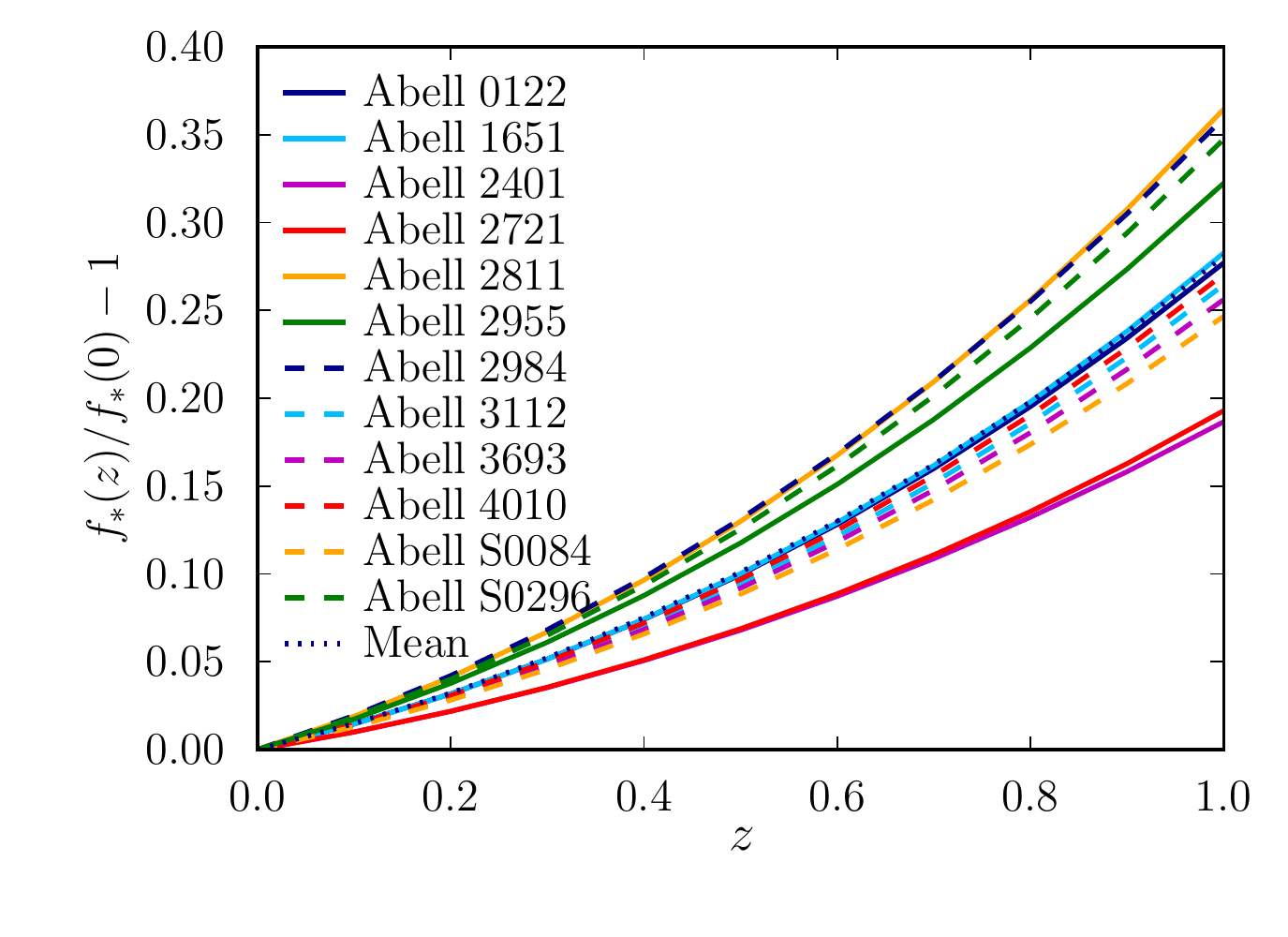}
\includegraphics[trim = 10mm 4mm 2mm 2mm, clip, scale=0.68]{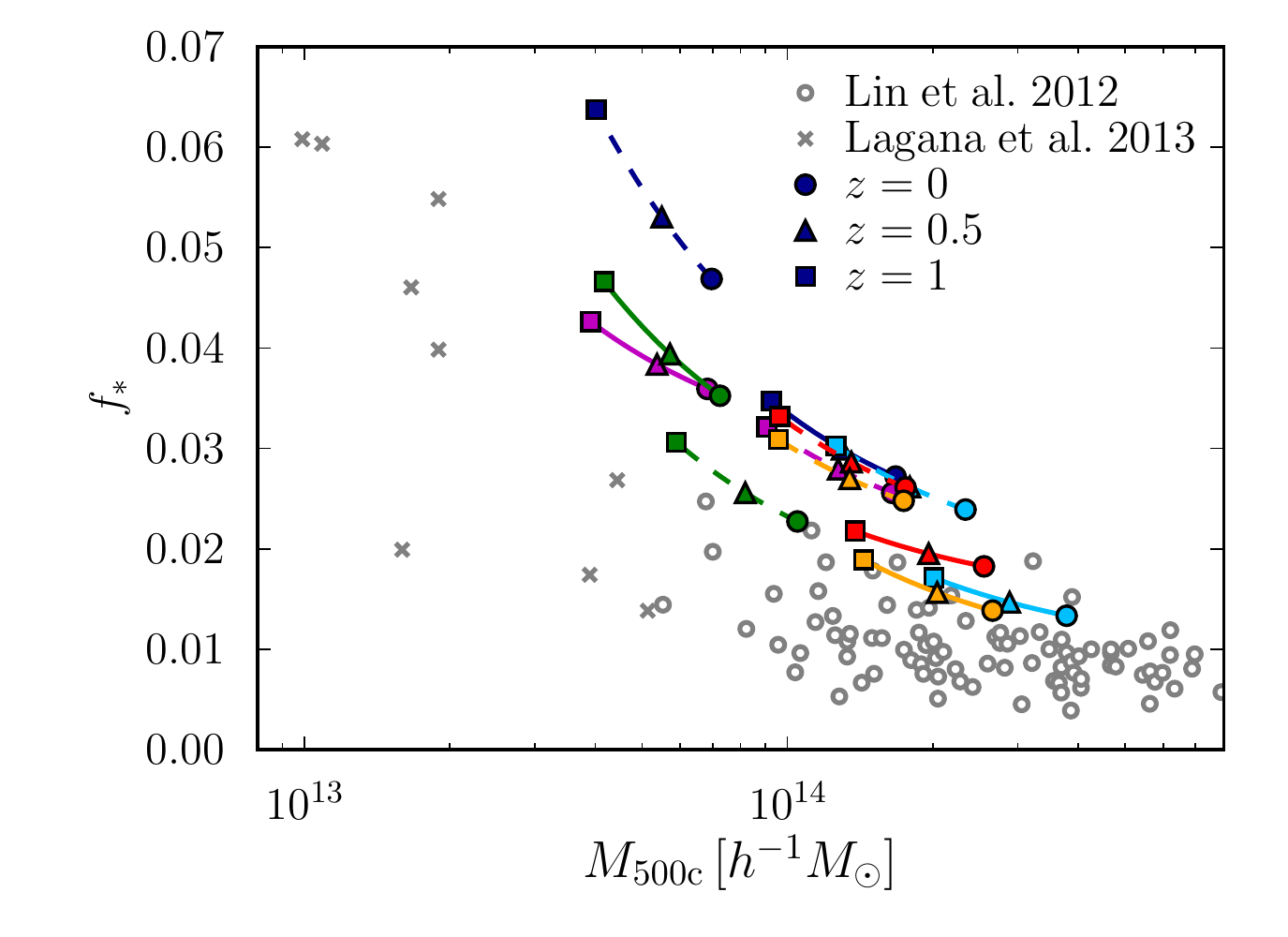}
\caption{Top panel: upper limits on the pseudo-evolution of the stellar mass fraction of 12 observed galaxy clusters \citep[][G13]{gonzalez_13}. Bottom panel: upper limits on the evolution of the same clusters on the $\mfoc - f_*$ plane. The circles, triangles and squares mark the values at $z = 0$, $0.5$ and $1$. For comparison, the observations of \citet{lin_12_baryoncontent} and \citet{lagana_13} (rescaled to match the stellar mass-to-light ratio and 3D deprojection used in G13) are plotted with gray circles and crosses. These stellar fractions are systematically lower than the values of G13 (see Figure 10 in G13), likely due to an unaccounted intra-cluster medium component. However, the figure shows that change in $f_*$ induced by pseudo-evolution occurs along the trend exhibited by the observed clusters.}
\label{fig:fstar}
\end{figure}

The results presented in the previous sections show that the \msr is not significantly affected by pseudo-evolution. Although we have only considered the \msr, some of our conclusions are applicable to other frequently used cluster scaling relations.

First, we note that the relation between the total mass within a given radius, $M(<r)$, and quantities that follow profiles similar to the total mass profile around the virial radius, such as the gas mass, $M_{\rm g}(<r)$, will not be significantly affected by pseudo-evolution, as both quantities will change by similar amounts. Second, we can conclude that the temperature-mass relation should be relatively unaffected by pseudo-evolution because the temperature profile is related to the mass profile via the hydrostatic equilibrium equation, qualitatively similar to the Jeans equation relating $\sigma$ and $M$. The slope of the temperature profile can thus be expected to be similar to the slope of the $\sigma(<r)$ profile. For the clusters analyzed in \citet{vikhlinin_06_clusters}, the temperatures averaged in the radial range of $0.15 - 1\rfoc$ have typical slopes of $\approx 0$ to $-0.2$ around $\rfoc$, similar to the slope of the velocity dispersion profile in the center panel of Figure~\ref{fig:profiles}. Therefore, the temperature-mass relation of both physically evolving and pseudo-evolving clusters should evolve according to the standard expectation arising from dimensional considerations \citep[][see also KB12 for a review]{kaiser_86_clusters},
\begin{equation}
T(<r) \propto \frac{M(<r)}{r} \propto \left[ \rhot^{1/2} M \right]^{\frac{2}{3}} \,.
\end{equation}
Given that both the $M_{\rm g}-M$ and $T(<r)-M$ relations are not expected to be significantly affected by pseudo-evolution, we also expect it to have little effect on the $Y$ parameter (KB12),
\begin{equation}
Y(<r) \propto M_{\rm g}(<r) \, T(<r) \,,
\end{equation}
related to the SZ flux. One should still keep in mind that pseudo-evolution can alter the fiducial evolution of the relation at the few-percent level, with the magnitude and sign of the effect depending on the specific choice of mass definition (see Section \ref{sec:interpretation:massdefinition} and Figures \ref{fig:analytical} and \ref{fig:msigevo2}). However, the small magnitude of the effect means that the pseudo-evolution can be parameterized as a small deviation around the fiducial evolution. 

We now turn to quantities whose radial profile may be very different from $M(<r)$, and from the $\sigma(<r)$ profile considered in this paper. In such cases, the effects of pseudo-evolution should be carefully evaluated. For example, the stellar mass profile can be quite different from the total mass profile because a significant fraction of the stellar mass is concentrated at the center, i.e. in the main cluster galaxy \citep[e.g.,][]{lin_04_icl, gonzalez_07_census}. One might thus expect that the stellar mass-total mass relation, $M_{\ast}(<R_{\Delta}) - M(<R_\Delta)$ \citep{lin_12_baryoncontent} could be significantly affected by pseudo-evolution. We consider this relation in some detail later in the paper, concluding that it is substantially affected by pseudo-evolution.

To estimate the effect of pseudo-evolution on the $M_{\ast}-M$ relation, we use the observed $M_{\ast}(<r)$ profiles of 12 clusters derived from XMM-Newton X-ray data \citep[hereafter G13]{gonzalez_13}. We model the central galaxy as a point mass, and both the stellar and total mass profiles as NFW profiles. We assign each cluster a concentration of the total mass profile according to the concentration-mass relation of \citet{zhao_09_mah}, giving values of $2.04 < c_{\rm 500c} < 2.59$. The concentration of the stellar mass profile was kept fixed at a value of $c_{\rm 500c} = 1.84$, corresponding to $c_{\rm 200c} = 2.9$ \citep{lin_04_lum}. The same NFW model was used in G13 to de-project the spatial distribution of galaxies. \citet{budzynski_12} found a slightly lower concentration of $c_{\rm 200c} = 2.6$, but their results are compatible with $c_{\rm 200c} = 2.9$ within errors. The exact value of the concentration has almost no impact on our results.

Assuming that both the stellar and total mass profiles of the clusters remain constant in physical units, and that $R_{500c}$ changes only as a result of pseudo-evolution, we find that the clusters exhibit an evolution of their stellar fraction, $f_{\ast} \equiv M_*(<\rfoc)/M_{\rm tot}(<\rfoc)$, shown in the top panel of Figure \ref{fig:fstar}. Given that the $M_{\rm tot}(<r)$ profile is steeper than the $M_{\ast}(<r)$ profile at radii around $\rfoc$, the increase in total mass as a result of pseudo-evolution is larger than the corresponding increase in stellar mass, thus decreasing the stellar fraction with time. Figure \ref{fig:fstar} shows that pseudo-evolution can lead to a decrease in the stellar fraction by up to $\approx 30\%$ between $z=1$ and $z=0$ for some clusters. The evolution is strongest for those clusters with the largest fraction of their stellar mass concentrated in the central galaxy, and thus a flatter overall $M_{\ast}(<r)$ profile. We note that almost no high-mass clusters are purely pseudo-evolving, meaning that the evolutions shown in Figure \ref{fig:fstar} represent upper limits on the real pseudo-evolution.

The bottom panel of Figure \ref{fig:fstar} shows the corresponding pseudo-evolution of clusters on the $f_{\ast}-\mfoc$ plane. Gray circles and crosses show the stellar fractions estimated by \citet{lin_12_baryoncontent} and \citet{lagana_13}. The individual clusters exhibit somewhat different pseudo-evolution tracks: the smallest mass clusters move along tracks of $f_{\ast} \propto \mfoc^{-0.5}$, while higher mass clusters evolve along somewhat flatter tracks. The latter, in fact, are quite close to the overall trend exhibited by observed clusters, $f_{\ast} \propto \mfoc^{-0.29}$ \citep{lin_12_baryoncontent}, i.e., pseudo-evolving halos evolve approximately along the observed relation. This co-evolution explains why no signature of pseudo-evolution was detected by \citet{lin_12_baryoncontent} who parameterized the relation and its evolution as
\begin{equation}
\label{eq:mstar_rel}
M_*(<\rfoc) \propto \mfoc^{\alpha} (1+z)^{\beta}
\end{equation}
and found $\alpha = 0.71 \pm 0.04$ and $\beta \approx -0.06\pm 0.22$, when clusters over a wide range of redshift are considered. It is tempting to interpret the observed relation as a consequence of pseudo-evolution, given the similarity of the pseudo-evolution tracks to the overall relation. This interpretation may seem unlikely, given that clusters are generally expected to be still evolving physically at low $z$. Nevertheless, \citet{wu_13_rhapsody1} report that that about a quarter of halos of masses as high as $M_{\rm vir}\approx 6\times 10^{14}\ M_{\odot}$ undergo pure pseudo-evolution at $z<1$. Thus, at masses $\lesssim 10^{14}\ M_{\odot}$, pseudo-evolution may well affect the majority of clusters and shape the observed $M_{\ast}-M$ relation. We note that this effect would be even more significant for the relation between the mass of a central galaxy and its total halo mass, because pseudo-evolution would result in tracks following $f_{\ast,\rm cen} \propto \mfoc^{-1}$, while the observed clusters exhibit $f_{\ast,\rm cen}\propto \mfoc^{-0.3\div 0.5}$.

Although further analyses need to be carried out to establish the exact role of pseudo-evolution in the evolution of the $M_{\ast}-M$ relation on cluster scales, it is clear that its effects are dominant for galaxy-scale halos in which most of the stellar mass is concentrated in the central galaxy (for example, $\approx 90\%$ of the Milky Way's stellar mass is concentrated in the central disk). We examine the impact of pseudo-evolution on the $M_{\ast}-M$ relation in a separate paper. 


\section{Conclusions}
\label{sec:conclusion}

Cluster-sized halos in a $\Lambda$CDM cosmology follow tight power-law scaling relations between their mass and observable properties. In this paper, we focus on the \msr, and analyze why the pseudo-evolution of halo radius and mass does not lead to obvious deviations from this relation. Our main conclusions are as follows.

\begin{enumerate}
\item For the $\mtoc$ mass definition, we reproduce the results of E08, and find that the slope and evolution of the relation roughly agree with the self-similar prediction $\sigma \propto \rhot^{1/6} M^{1/3}$. Furthermore, we find only very small differences between the evolution of the relations for physically accreting and pseudo-evolving halos, even though they follow very different evolutionary tracks on the \ms plane.

\item We find that the similar evolution arises because all halos exhibit a tight relation between their $\sigma(<r)$ and $M(<r)$ profiles, such that  $\sp=\sigma(<r)/\sqrt{GM(<r)/r}$ is approximately constant at $r\approx [R_{2500c}-R_{200m}]$.  Thus, as the outer  radius, $R$,  increases as a result of pseudo-evolution, the halos approximately preserve their \msr. We show that such a relation is generically expected for a wide range of halo profiles under the condition of Jeans equilibrium. 

\item This result highlights the fact that tight scaling relations are the result of tight relations between radial profiles of physical quantities, which explains why such relations are almost insensitive to the choice of the virial radius definition. Exceptions are at very small, $r\lesssim R_{2500c}$, and very large radii, $r\gtrsim R_{200m}$, where we find small deviations of the profiles from the relations they exhibit at intermediate radii, and show that their \msr deviates from the predicted redshift scaling by a few percent as a result of pseudo-evolution.
  
\item We discuss the implications of these results for other cluster scaling relations and argue that pseudo-evolution should have very small effects on most scaling relations, except for those relations that involve the stellar masses of galaxies. In particular, we show that the relation between the stellar mass fraction and total mass is affected by pseudo-evolution and is largely shaped by it for halo masses $\lesssim 10^{14}\ M_{\odot}$.

\end{enumerate}

Our results show that pseudo-evolution should not appreciably affect the evolution of most scaling relations, such as $Y-M$ and $T-M$, and their scatter. At the same time, the stellar--halo mass relation is significantly affected by pseudo-evolution at $\lesssim 10^{14}\ M_{\odot}$, which warrants further investigation. We will explore these effects on galaxy mass scales in a future paper. 


\section*{Acknowledgments}

We thank Matt Becker for his help with running the Gadget2 simulation, the halo finding and extracting the density profiles. We are also grateful to Alexey Vikhlinin and Eduardo Rozo for useful discussions. We thank Douglas Watson, Douglas Rudd and Matt Becker for their comments on the draft. We have used computing resources provided by the University of Chicago Research Computing Center to carry out the simulation used in this study. AVK acknowledges support by the NSF and NASA via grants OCI-0904482 and NNX13AG81G and by the Kavli Institute for Cosmological Physics at the University of Chicago through grants NSF PHY-0551142 and PHY-1125897 and an endowment from the Kavli Foundation and its founder Fred Kavli. 


\bibliographystyle{apj}
\bibliography{../../../_LatexInclude/sf.bib}

\end{document}